\documentclass[compsoc,journal]{IEEEtran}

\usepackage{hyperref}
\usepackage{xcolor,colortbl}
\newcommand{\fig}[1]{Figure~\ref{fig:#1}}

\newcommand{\tion}[1]{\S\ref{sec:#1}}
\usepackage{balance}
\usepackage{wrapfig}
\usepackage{multirow}
\usepackage{blindtext, graphicx}
\usepackage{boldline}
\usepackage{subfig}
\usepackage{booktabs}
\usepackage{ragged2e}
\usepackage{dblfloatfix}
\usepackage[para,online,flushleft]{threeparttable}
\newcommand{\IT}{HARMLESS}

\usepackage{lipsum}

\usepackage[tikz]{bclogo}
\newenvironment{RQ}[1]%
{\noindent\begin{minipage}[c]{\linewidth}%
\begin{bclogo}[couleur=gray!25,%
                arrondi=0.1,%
                logo=\bctrombone,%
                ombre=true]{{\small ~#1}}}%
{\end{bclogo}\end{minipage}\vspace{2mm}}

\newcommand{\bi}{\begin{itemize}}
\newcommand{\ei}{\end{itemize}}

\usepackage{enumitem}
\setlist[itemize]{leftmargin=*}
\setlist[enumerate]{leftmargin=*}
\usepackage{makecell}
\usepackage[linesnumbered,ruled,vlined]{algorithm2e}
\setlist{nolistsep}

\setlist[1]{itemsep=0pt}

\usepackage{amsmath}

\usepackage{color}
\usepackage{listings}
\usepackage{times}
\usepackage{colortbl}
\usepackage{tikz}
\def\checkmark{\tikz\fill[scale=0.4](0,.35) -- (.25,0) -- (1,.7) -- (.25,.15) -- cycle;}

\makeatletter
\let\th@plain\relax
\makeatother

\definecolor{Gray}{rgb}{0.88,1,1}
\definecolor{Gray}{gray}{0.85}
\definecolor{lightgray}{gray}{0.8}
\usepackage[framed]{ntheorem}
\usepackage{framed}
\usepackage{tikz}
\usetikzlibrary{shadows}
\theoremclass{Lesson}
\theoremstyle{break}

\tikzstyle{thmbox} = [rectangle, rounded corners, draw=black,
fill=gray!25,  drop shadow={fill=black, opacity=1}]

\newshadedtheorem{lesson}{Answer to RQ}

\def\checkmark{\tikz\fill[scale=0.4](0,.35) -- (.25,0) -- (1,.7) -- (.25,.15) -- cycle;}

\definecolor{myGray}{RGB}{220, 220, 220}
\definecolor{myRed}{RGB}{156, 4, 4}
\definecolor{myBlue}{RGB}{108, 156, 236}
\definecolor{myGreen}{RGB}{108, 172, 76}

\usepackage{xcolor}

\bstctlcite{IEEEexample:BSTcontrol}

\begin{document}

\SetKwProg{Fn}{Function}{}{}
%

\author{Zhe Yu,~
        Christopher Theisen,~
        Laurie Williams,~\IEEEmembership{Fellow,~IEEE},
        and Tim Menzies,~\IEEEmembership{Fellow,~IEEE}
\IEEEcompsocitemizethanks{\IEEEcompsocthanksitem Z. Yu, L. Williams, and T. Menzies are with the Department
of Computer Science, North Carolina State University, Raleigh, USA.\protect\\
E-mail: \{zyu9, lawilli3\}@ncsu.edu, timm@ieee.org.
\IEEEcompsocthanksitem C. Theisen is with Microsoft, Seattle, USA.\protect\\
E-mail: theisen.cr@gmail.com}}




\title{Improving Vulnerability Inspection\\ Efficiency Using Active Learning  }

\IEEEtitleabstractindextext{%
{\justify\begin{abstract}
Software engineers can find vulnerabilities with less effort if they are directed towards   code that might contain more  vulnerabilities.  
{\IT} is an incremental support vector machine tool that builds
a vulnerability prediction model from the source code   inspected to date, then suggests what   source code files  should   be inspected next.
In this way, {\IT} can  reduce the time and effort required to achieve some desired level of recall
for finding vulnerabilities. The tool also provides feedback on when to  stop (at that desired level of recall)
while at the same time,  correcting human errors by double-checking suspicious files.

This paper evaluates {\IT} on Mozilla Firefox vulnerability data.
{\IT} found 80, 90, 95, 99\% of the vulnerabilities by inspecting 10, 16, 20, 34\% of the source code files. 
When targeting 90, 95, 99\% recall, {\IT} could stop after inspecting 23, 30, 47\% of the source code files.
Even when human reviewers fail to identify half of the vulnerabilities (50\% false negative rate), {\IT} could detect 96\% of the missing vulnerabilities by double-checking half of the inspected files.

Our results serve to highlight the very steep cost of protecting software from vulnerabilities (in our case study that cost is, for example, the  human effort of inspecting 28,750 $\times$ 20\% = 5,750 source code files to identify 95\% of the vulnerabilities).  While
this result could benefit the mission-critical projects where human resources are available for inspecting thousands of source code files, the research challenge for future work is how to further reduce that cost. The conclusion of this paper discusses various ways that goal might be achieved.


\end{abstract}}
}

\IEEEdisplaynontitleabstractindextext

\maketitle
\section{Introduction}
\label{sec:introduction}


Software security is a current and urgent issue.  A recent report~\cite{black2016dramatically} from the US National Institute of Standards and Technology (NIST) warns that ``current systems perform increasingly vital tasks and are widely known to possess {\em vulnerabilities}''. The vulnerabilities discussed in this paper are defined
as follows:
\bi
\item A mistake in software that can be directly used by a hacker to gain access to a system or network; or
\item A mistake that lets  attackers  violate a security policy~\cite{lewiscommon}.
\ei
Government and scientific bodies stress the need for reducing software vulnerabilities. In a report to the White House Office of Science and Technology Policy, ``Dramatically Reducing Software Vulnerabilities''~\cite{black2016dramatically}, NIST encourages more research on approaches to reduce security vulnerabilities.  The need to reduce vulnerabilities is also emphasized in the 2016 US Federal Cybersecurity Research and Development Strategic Plan~\cite{trautman2015cybersecurity}.


To  protect against software vulnerabilities, teams need to detect and fix the vulnerabilities before deployment. In this paper, we focus on the ``detect'' part of this process. Code inspection is a key process for vulnerability detection. Such inspections  require software engineers to inspect large amounts of code (e.g. check that no call to the ``C'' {\tt printf} function can be supplied a format string that is mismatched to the type of the items being printed). 
Resource limitations often preclude software engineers to inspect all source code files~\cite{slaughter1998evaluating}. Making informed decisions on what source code files to inspect can improve a team's ability to find vulnerabilities.
Vulnerability prediction models (VPMs) make such informed decisions by learning a machine learning model from known vulnerabilities. The model is then applied to classify source code files as ``vulnerable'' or ``non-vulnerable''. If software engineers only inspect the predicted ``vulnerable'' files,  then human effort is reduced. That is, a good VPM helps human to find {\em more} vulnerabilities by inspecting {\em less} code.

Although the state-of-the-art VPMs have shown promising results on finding vulnerabilities with reduced human inspection effort~\cite{scandariato2014predicting,shin2013can,zimmermann2010searching,Theisen2015ICSE,theisen2017software}, these approaches have limitations:
\bi
\item
VPMs learn models using training
data which describes known vulnerabilities. Prior to the initial release, such data may not be available\footnote{Transferring training data may not solve this problem. Cross project vulnerability prediction has been shown to perform worse than within project vulnerability prediction~\cite{walden2014predicting}.}.
\item
Existing VPMs do not allow users to choose what level of recall (percentage of vulnerabilities found) to reach. Moreover, when software engineers finish inspecting the selected source code files, they do not know how many more vulnerabilities are as-yet-undetected in the remaining code.
\item
Humans may inspect a  file but fail to find  vulnerabilities~\cite{hatton2008testing}. Double 
checking (where files are inspected by other humans) is needed to cover such  errors. Current VPM research does not discuss how to design  cost-effective double-checking strategies.
\ei

\noindent
The central insight of this paper is that the vulnerability prediction problem (finding more vulnerabilities by inspecting less code) belongs to a class of information retrieval problem---the  {\em total recall}~\cite{grossman2016trec} problem
(described in \tion{related_work:Total Recall}). An active learning-based framework has been shown effective to resolve all the above-mentioned limitation for total recall problems in another domain---selecting relevant papers
for literature reviews~\cite{Yu2019}. 
This paper checks the conjecture that this active learning-based framework for selecting relevant papers can be applied and adapted to help find vulnerabilities efficiently while
mitigating the limitations of current vulnerability prediction approaches.
That said, in this paper, a novel direction is presented to assist software engineers in identifying vulnerabilities efficiently. A vulnerability inspection tool {\IT} is built by applying and adapting the active learning-based framework for selecting relevant papers.  Using {\IT}, engineers inspect some source code files while {\IT} trains/updates a VPM based on the inspection results. Then {\IT} applies the VPM to suggest which files should be inspected next. Through iterating this process, {\IT} guides the human inspection efforts towards source code files that are more likely to contain vulnerabilities. An interesting finding here is that the active learning-based framework can be applied to vulnerability prediction problem (as {\IT}) with barely any modification. This means two things---firstly we can improve vulnerability inspection efficiency by recasting the problem in terms of total recall. Secondly, in the future, innovations in the solution of any total recall problem, including vulnerability prediction, can improve all other total recall problems including vulnerability prediction. This is an exciting line of research since this suggests a  synthesis of many (seemingly different) lines of research. 

This paper assesses the effectiveness of {\IT}  by simulating code inspections on C and C++ files from Mozilla Firefox project. The authors tagged the known vulnerabilities of the Mozilla Firefox project from Mozilla Foundation Security Advisories blog~\cite{mozillablog} and bug reports from Bugzilla\footnote{https://bugzilla.mozilla.org/} up to November 21st, 2017. Among the 28,750 unique source code files, 271 files contain vulnerabilities. These known vulnerable files are treated as ground truth. Note that our dataset does not provide information on whether an actual incident was caused by the identified ground truth vulnerabilities. During our file-level simulations, when a human is asked to inspect a source code file and tell whether it contains vulnerabilities or not, the ground truth is applied instead of a real code inspection. This enables our experiments to be repeated multiple times with different algorithm setups and provides reproducibility of this paper. For full details on that data, see \tion{case-study}.

Using this data, we explore four research questions to see whether {\IT} can better resolve the aforementioned limitations of traditional VPMs:

 {\bf RQ1: Can human inspection effort be saved by applying {\IT} to find a certain percentage of vulnerabilities?}
Simulated on the Mozilla Firefox data without prior known vulnerabilities as training data, we show that 60, 70, 80, 90, 95, 99\% of the known vulnerabilities can be found by inspecting around 6, 8, 10, 16, 20, 34\% of the source code files, respectively. These results show that a good amount of human effort can be saved by applying {\IT}. 
This result also highlights the very
steep costs associated with protecting   software
from vulnerabilities.
Even with our best methods, for the Mozilla Firefox dataset studied here, humans still need  to  manually inspect 28,750 $\times$ 20\% = 5,750 source code files to identify 95\% of the vulnerabilities.
 While this result could benefit the mission-critical projects where human resources are available for inspecting thousands of source code files (as done in~\cite{zelkowitz2001understanding}), the clear research challenge for future work is how to further reduce that cost. The conclusion
 of this paper offers some notes
 on how that goal might be achieved.

Note that these {\bf RQ1} results  assume that humans are infallible; i.e. they make no mistakes in their inspections (see {\bf RQ3, RQ4} for what happens when humans become fallible).

{\bf RQ2: Can {\IT} correctly stop the vulnerability inspection when a predetermined percentage of vulnerabilities has been found?}

We show that {\IT} can accurately
estimate the number of remaining
vulnerable files in a project. Based on the estimation, {\IT} can tell humans when they should stop the inspection
(i.e. when they have found the predetermined percentage of vulnerabilities). Our results also suggest that {\IT} tends to slightly over-estimate the number of remaining vulnerable files, e.g. when targeting 95\% recall, the inspection stopped at 97\% recall with 30\% cost, which means 97\% of the vulnerable files can be identified by inspecting 8,625 source code files. Compared to the result of RQ1, about 3,000 more files need to be inspected due to the small estimation error of {\IT}. 
 Hence, it is very important to further improve the estimation since a tiny reduction in its error can lead to the saving of a large amount of effort.

{\bf RQ3: Is {\IT}'s performance affected by human errors?}
Hatton~\cite{hatton2008testing} warns that when  inspecting codes for defects, human fails to detect 47\% of the defects (47\% false negative rate and 0\% false positive rate) in average. We assume the same error rate for vulnerability inspections. By adding in 10 to 50\% false negative rate to the simulated human oracles randomly, we simulate the influence of growing human error rate on {\IT}'s performance. Such human errors drastically and negatively impact the inspection result, thus motivating the next research question.

{\bf RQ4: Can {\IT} correct human errors effectively?}
As seen in our simulations in \tion{errors}, {\IT} outperforms the state-of-the-art error correction mechanisms by only double-checking 50\% of the inspected files but covering 96\% of the missing vulnerabilities. According to our results, when targeting 95\% recall with human having 50\% chance of missing a vulnerability during the inspection, {\IT} can reach $0.95\times0.85 = 81\%$ recall for $0.21\times1.86 = 39\%$ cost, which means 81\% of the vulnerable files can be identified by inspecting 11,230 files (including double checks). If the reader finds this to be an excessive amount, then they might want to consider what has to happen without our tools: 
\bi
\item
Without {\IT},
one engineer would only find 50\% of the vulnerabilities, and
only after inspecting 28,750 files (100\% of the files); 
\item
Without {\IT},
two engineers would  find 75\% of  the vulnerabilities,
but
only
after 
inspecting 57,500 files (200\% of the files including double-checking effort).
\ei

\subsection{Contributions of this Paper}
\begin{enumerate}
\item
Here in this paper we present a different direction to assist software engineers identifying potential vulnerabilities efficiently. Instead of training a model to predict vulnerabilities like traditional VPMs do, we apply an active learning-based framework (\IT) to learn from human inspection results and to better select which source code files should be inspected next.
\item
Simulations on Mozilla Firefox vulnerability dataset show that, without any prior knowledge, vulnerability inspections with {\IT} can identify most vulnerabilities by inspecting only a small portion of the codebase.
\item
An estimator that can be applied along with the execution of {\IT} to estimate the remaining number of vulnerabilities in the codebase. This estimator can be used to provide guidance to stop the vulnerability inspection when a desired percentage of vulnerabilities have been found.
\item
An error correction mechanism that detects vulnerabilities that humans failed to detect without imposing excessive extra human inspection effort.
\item
All code and data used in this analysis
are available\footnote{\url{https://github.com/ai-se/Mozilla_Firefox_Vulnerability_Data}}, allowing other researchers to replicate, improve, or even refute our findings.
\end{enumerate}

The rest of this paper is structured as follows. Some background and related work is discussed in \tion{related-work}. Our methodology is described in  \tion{methodology}. This is followed by the details on how to simulate the {\IT} inspection process on Mozilla Firefox vulnerability dataset in \tion{case-study}. Details of the experiment (simulation) designs and answers to the research questions are presented in \tion{experiments}.  Threats to validity and limitations to this work are analyzed in \tion{discussion} while conclusion and future work are provided in \tion{conclusion}.

Before beginning, we digress to comment
that in this case study we do {\em not} show that in practice, tools like {\IT} are effective in
reducing vulnerabilities in real projects. Before we can ask projects to work with us on such a study, we must first certify our technique on carefully controlled laboratory problems (hence this paper).

\section{Background and Related Work}
\label{sec:related-work}

\subsection{Security Vulnerability Prediction and Prioritization}\label{sec:related-work:vuls}

In this subsection, we discuss the existing vulnerability prediction models (VPMs). These prediction models are used to prioritize validation and verification efforts. Based on what information is required and utilized in those approaches, we categorize them into three groups: 
\begin{enumerate}
\item
{\em Supervised methods}, which trains a model on historical inspection results (known vulnerable files and known non-vulnerable files) to predict the likelihood of new/changed files containing vulnerabilities; 
\item
{\em Semi-supervised methods}, which trains a model on historical inspection results as well as the new/changed (unlabeled) files to predict the likelihood of new/changed files containing vulnerabilities; 
\item
{\em Unsupervised methods}, which do not rely on historical inspection results, but utilizes other indicators for vulnerability proneness. 
\end{enumerate}
\subsubsection{Supervised Methods}
Most of the previous work in vulnerability prediction are supervised; 
i.e. they use  known vulnerabilities to train a supervised learning model for predicting unknown vulnerabilities. Focusing on statistical classifiers on source code artifacts, such as binaries, files, and functions, these approaches classify code artifacts as either vulnerable or not vulnerable.  Some of the datasets used in this process include the Windows operating system~\cite{zimmermann2010searching,Theisen2015ICSE}, the Mozilla Firefox web browser~\cite{shin2011evaluating,theisen2017software}, and the RedHat Enterprise Linux Kernel~\cite{shin2011evaluating}. The two types of features they use for predicting vulnerabilities are software metrics and text mining features:

\textbf{Software metrics:} Zimmermann {\it et al.}~\cite{zimmermann2010searching} reported a median recall of 20\% 
after  using all available source code metrics  to build a Random Forest~\cite{breiman2001random} classifier. Similarly, Chowdhury {\it et al.} ~\cite{chowdhury2011using} report that their vulnerability predictors had low recalls of 23\%. Shin {\it et al.}~\cite{shin2008complexity,shin2011evaluating,shin2013can} focused on churn and complexity with a different set of software metrics and achieved a higher recall of 83\%. However, they also reported a precision as low as 5\%. Hovsepyan {\it et al.}~\cite{hovsepyan2014design} extracted software metrics from design churn and achieved 71\% precision, but only 27\% recall on average.

\textbf{Text mining features:} Scandariato {\it et al.}~\cite{scandariato2014predicting} applied
text mining to predict security vulnerabilities. Their hypothesis was  that specific tokens found in source code files may indicate whether this file is more prone to vulnerabilities; for example, the presence of ``nullptr'' might mean a specific source code file is more prone to vulnerabilities resulting from null references. In their case studies~\cite{scandariato2014predicting}, a static code analysis tool was used to decide the ground truth labels. Such static code analysis tools have a notoriously large false positive rate and may incorrectly decide that many code components are ``vulnerable''. Neuhaus {\it et al.}~\cite{neuhaus2007predicting} extracted only the imports and calls information from source code as text mining features to predict vulnerabilities. They reported that their approach correctly predicted about half of all vulnerable components, and about two-thirds of all
predictions were correct. Walden {\it et al.}~\cite{walden2014predicting} compared software metric-based approaches for vulnerability prediction to text mining approaches and found that text mining performed better in terms of precision and recall than software metric approaches. Later in \tion{experiments}, we show the same trend with {\IT} that text mining features outperform software metrics. Walden {\it et al.}~\cite{walden2014predicting} also reported that cross-project prediction performances were generally poor---49\%, 36\% of the source code files need to be reviewed for finding 66\%, 70\% of the vulnerabilities, respectively. These results suggest that training data from the same project are required for these VPMs. As a result, 
these supervised VPM methods cannot be used before the first release since no training data from the same project is available.

\subsubsection{Semi-supervised Methods}

Semi-supervised VPMs not only learn from the known vulnerable and non-vulnerable files, but also 
extrapolate their conclusions
to files that  have not been inspected yet~\cite{zhu2005semi,chapelle2009semi,le2016budgeted,gieseke2014fast}. For example, Meng {\it et al.}~\cite{7853039} (a)~applied a label propagation algorithm~\cite{zhu2005semi} to assign labels to the unlabeled data; then (b)~used those guesses
to train a   model for buffer overflow prediction. 
To the best of our knowledge, Meng's work is the only prior work on vulnerability prediction that used semi-supervised learning. Note that their work only predicts for   one type of security vulnerabilities (buffer overflow) while in our work, we can target at any type of  vulnerabilities.

\subsubsection{Unsupervised Methods}

Unsupervised approaches do not rely on human oracles (of which files contain vulnerabilities) to build the VPM. Instead, they are fully automatic methods that rely on other indicators of vulnerability proneness. 
For example, Gegick {\it et al.}~\cite{gegick2009toward} trained a decision tree classifier on non-security failures to rank software components. They found that, 57\% of the
vulnerable components were in the top 9\% of the total component
ranking, but with a 48\% false positive rate.  A better result was achieved by Theisen {\it et al.}~\cite{theisen2015strengthening,Theisen2015ICSE,theisen2017risk} using crash features extracted from crash dump stack traces, which are collected after the software's first release. The crash features were used to approximate the attack surface and predict which parts of the software are more prone to be vulnerable. Theisen {\it et al.} found that crash history is a strong indicator of vulnerabilities---48.4\% of the ``crashed'' binaries in Windows contain 94.6\% of known vulnerabilities~\cite{Theisen2015ICSE}, and 15.8\% of the ``crashed'' source code files in Mozilla Firefox contain 73.6\% of known vulnerabilities~\cite{theisen2017risk}. The advantage of this approach is that it does not require the labeling of training data. However, one major drawback of Theisen {\it et al.} approach is that higher recall cannot be achieved since no information is provided in those source code files without crash history.



\subsection{Active Learning}
\label{sec:related_work:Total Recall}

The methods listed above represent some of the
best VPM results seen to date. While they are all
significant research results, there is much
remaining to be done. In this paper, we see
if incrementally adding human insights (via active
learning) to supervised/semi-supervised methods can mitigate any of the shortcomings discussed in \tion{introduction}.

\begin{wrapfigure}{r}{1.6in}
\includegraphics[width=1.55in]{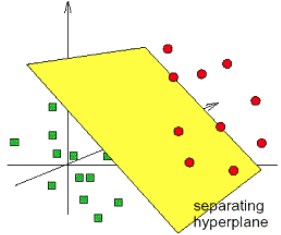}
\caption{Separating positive (red circles) from  negative (green squares) examples.}\label{fig:svm}
\end{wrapfigure}
The key idea behind active learning is that a machine learning algorithm can train  faster (i.e. using less  data) if it is allowed to choose the data from which it learns~\cite{settles2012active}. 
To understand active learning, consider the decision plane between the positive and negative examples in \fig{svm}. Suppose we want to find more positive examples and we have access to the  \fig{svm} model. One tactic for quickly finding those positive examples would be to inspect unlabeled data that fall into the red region of this figure, as far as possible from the green ones (this tactic is called {\em certainty sampling}). Another tactic would be to better improve the current model by verifying the position of the boundary; i.e. inspecting unlabeled data that are closest to the boundary (this tactic is called {\em uncertainty sampling}). The experience from explored total recall problems is that  such {\em active learners} outperform supervised and semi-supervised learners and can significantly reduce the effort required to achieve high
recall~\cite{Cormack2017Navigating,Cormack2016Engineering,cormack2016scalability,cormack2015autonomy,cormack2014evaluation,grossman2013,wallace2010semi,wallace2010active,wallace2011should,wallace2012class,wallace2013active,Yu2018,Yu2019}.


For example,
in electronic discovery, attorneys are hired to review a large number of documents looking for relevant ones to a certain legal case and provide those as evidence. Cormack and Grossman~\cite{cormack2015autonomy,cormack2014evaluation,grossman2013} designed and applied continues active learning to save attorneys' effort from reviewing non-relevant documents, which further can save a large amount of the cost of legal cases. Cormack and Grossman also proposed ``knee methods''\footnote{Where performance is monitored until
the performance growth curve ``turns a corner''.}
as stopping rule~\cite{Cormack2016Engineering} and as a way of efficiently correcting human errors~\cite{Cormack2017Navigating}.

Also, in  evidence-based medicine, researchers screen titles and abstracts to determine whether one paper should be included in a certain systematic review. Wallace {\it et al.}~\cite{wallace2010semi} designed patient active learning to help researchers prioritize their effort on papers to be included. With patient active learning, half of the screening effort can be saved while still including most relevant papers~\cite{wallace2010semi}. Similarly, Miwa {\it et al.}~\cite{miwa2014reducing} used active learning but with certainty sampling and weighting to achieve better efficiency. In addition, Wallace {\it et al.}~\cite{wallace2013active} tried to estimate the number of relevant papers to indicate what recall has been reached during the process.

\section{Methodology}\label{sec:methodology}

The central insight of this paper is that the vulnerability prediction problem (finding more vulnerabilities by inspecting less code) along with the electronic discovery and evidence-based medicine problems  all belong to a class of information retrieval problem called the {\em total recall} problem~\cite{grossman2016trec}.

\subsection{Total Recall}
\label{sec:total-recall}

The target of {\em total recall} is to optimize the cost for achieving very high recall (ideally, very close to 100\%) with a human assessor in the loop~\cite{grossman2016trec}. It differs from ad hoc information retrieval where the objective is to identify the best, rather than all relevant information, and from classification or categorization where the objective is to separate relevant from non-relevant information based on previously labeled training examples~\cite{cormack2018quest}. More specifically, the total recall problem can be described as follows:
\begin{RQ}{The Total Recall Problem:}
Given  candidates $E$ with a small positive fraction $R \subset E$, each  $x \in E$ can be inspected to reveal its label as positive ($x\in R$) or negative ($x \not\in R$) at a cost. Starting with the labeled set $L = \emptyset$, the task is to inspect and label as few candidates as possible ($\min |L|$) while achieving very high recall in finding positives ($\max |L\cap R|/|R|$).
\end{RQ}
The core problem of vulnerability prediction is how to find most of the vulnerabilities with least code inspected. Therefore the vulnerability prediction problem can be generalized as the total recall problem with:
\bi
\item 
$E$: the entire codebase of a software project.
\item
$R$: set of source code files that contain at least one target type vulnerability.
\item
$L$: set of source code files already inspected by humans.
\item
$L_R = L\cap R$: set of source code files already inspected by humans and contain at least one target type vulnerability.
\ei
In addition, we believe it is even more appropriate to treat the vulnerability prediction problem as a total recall problem rather than a classification problem, given the severe consequences missing vulnerabilities might have. Therefore we conjecture that the active learning-based framework for total recall problems can  better address the vulnerability prediction problem. The rest of this paper checks this conjecture.

\subsection{Active learning-based Framework}
\label{sec:total_recall_solution}

\begin{figure}[!th]
    \centering
    \includegraphics[width=\linewidth]{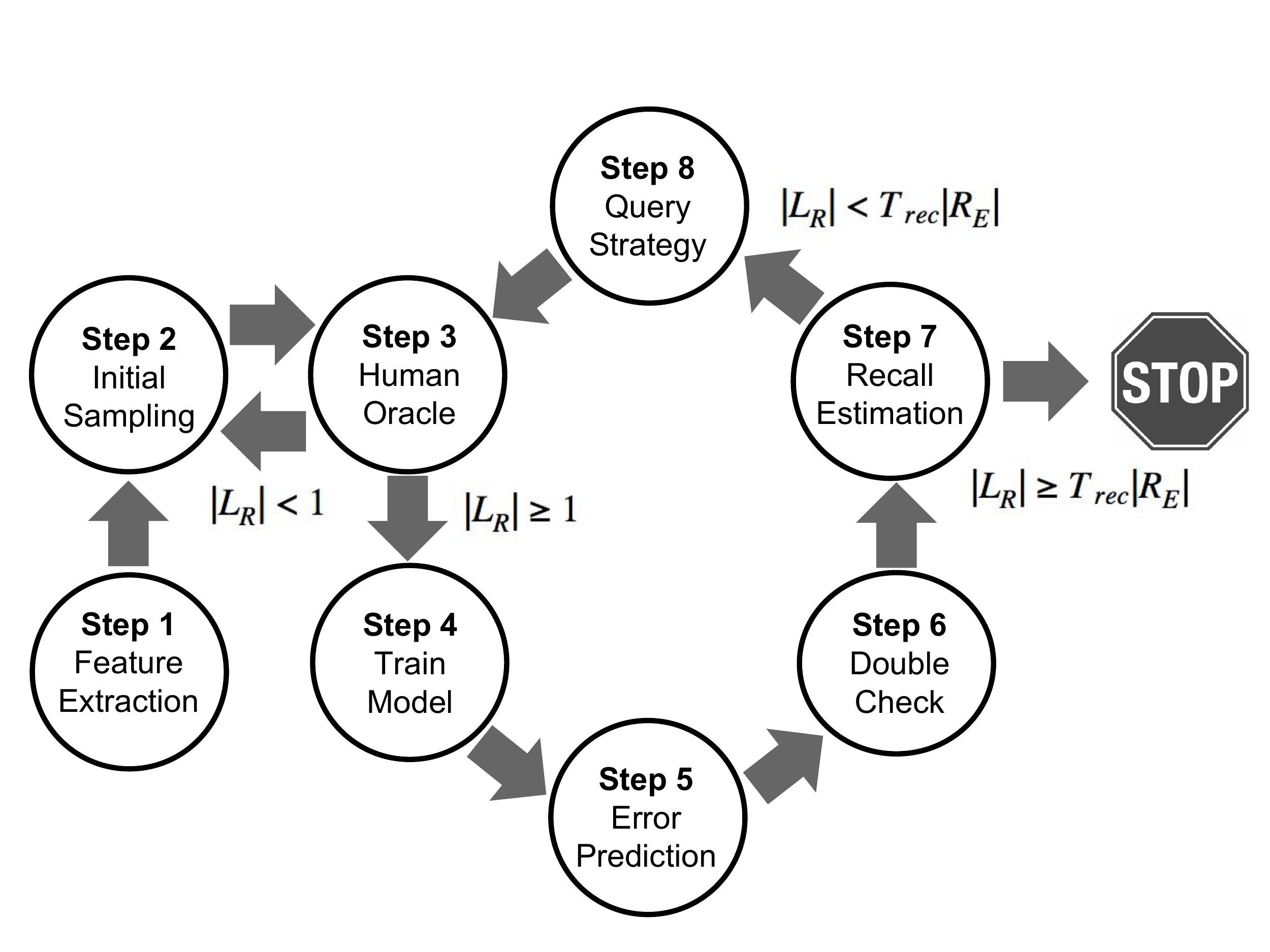}
    \caption{Block diagram of the active learning-based framework for total recall problems}
    \label{fig:{\IT}}
\end{figure}

In practice, active learning-based solutions of total recall problems focus on the following three targets~\cite{wallace2013active,Cormack2016Engineering,Cormack2017Navigating,Yu2019}:
\bi
\item
\textbf{Target 1 efficiency}: achieving higher recall with a lower cost than other solutions. This is the main target of total recall problems as defined in \tion{total-recall}. This target measures the performance of a total recall solution with the true recall $|L_R|/|R|$ in an ``ideal'' scenario when every inspected examples are correctly labeled. 
\item
\textbf{Target 2 stopping rule}: stopping at a predetermined recall rather than an arbitrary stopping point varying from different projects. Given that the total number of positives $|R|$ is unknown before all the candidates $E$ are inspected, stopping at a predetermined recall is especially hard~\cite{wallace2013active}. Stopping too late would waste inspection effort while stopping too early would miss potential positives (e.g. vulnerabilities).
\item
\textbf{Target 3 human error correction}: correcting human errors efficiently. Given the ``human-in-the-loop'' nature of the total recall problems, human errors affect the inspection results a lot. How to design strategies to efficiently fix such human errors is also an important problem.
\ei

In our previous work~\cite{Yu2018}, we refactored techniques from Cormack, Wallace, and Miwa {\it et al.}~\cite{cormack2014evaluation,wallace2010semi,miwa2014reducing} and created an active learning-based framework to solve one specific total recall problem (identifying relevant papers for literature reviews while minimizing the number of candidate literature being reviewed by humans). 
In that work, an SVM model incrementally learns which papers are more likely to be ``relevant'' based on previous decisions from humans and then guides the humans to read those papers next. When tested on four datasets, the proposed active learning framework outperforms the prior state-of-the-art approaches in terms of \textbf{Target 1 efficiency}~\cite{Yu2018}. Furthermore, SEMI, an estimator for the total number of relevant papers and DISAGREE, a technique to efficiently correct human classification errors were developed as part of the framework for \textbf{Target 2 stopping rule} and \textbf{Target 3 human error correction}, respectively~\cite{Yu2019}.  When combined, all these techniques result in a general active learning-based framework shown in Figure~\ref{fig:{\IT}} where \textbf{Step 4} and \textbf{Step 8} focus on \textbf{Target 1 efficiency}, \textbf{Step 7} decides \textbf{Target 2 stopping rule}, and \textbf{Step 5} and \textbf{Step 6} resolve \textbf{Target 3 human error correction}. The framework helps humans retrieve a target percentage ($T_{rec}$) of relevant information ($R$) with reduced human effort ($L$) as follows:

\begin{enumerate}[start=1,label={\bfseries Step \arabic*}]
\item
\label{step:1}
\textbf{Feature Extraction:} Given a set of candidates $E$, extract features from each candidate. Initialize the set of labeled data points as $L \leftarrow \emptyset$ and the set of labeled positive data points as $L_R \leftarrow \emptyset$.
\item
\label{step:2}
\textbf{Initial Sampling: }Sample without replacement $N_1$ unlabeled data points to generate a queue $Q\subset (E \setminus L)$~\footnote{Let $A$ and $B$ be two sets.  The set difference, denoted $A\setminus B$ consists of all elements of $A$ except those which are also elements of $B$.}.
\item
\label{step:3}
\textbf{Human Oracle: }Seek human oracles for data points in the queue $x \in Q$: label $x$ as positive or negative after a human inspects the source code file; add file $x$ into the labeled set $L$ and remove $x$ from $Q$; if $x$ is labeled positive, also add $x$ into set $L_R$.  Once the queue is cleared $Q=\emptyset$, proceed to \ref{step:4} if at least one positive data point has been found ($|L_R|\ge 1$), otherwise go back to \ref{step:2}.
\item
\label{step:4}
\textbf{Train Model: }Train a classifier on the current labeled set $L_R$ and $L$ with features extracted in \ref{step:1} to predict whether one data point is positive or negative.
\item
\label{step:5}
\textbf{Error Prediction: } Apply the classifier trained in \ref{step:4} to predict on the labeled set $L$. Select $N_2$ data points with highest predicted probability to have been wrongly labeled for double-checking.
\textit{This step aims at correcting human errors more efficiently by suggesting which data points are more likely to have been wrongly labeled.}
\item
\label{step:6}
\textbf{Double check: }Each file in the queue from \ref{step:5} is assigned to different humans for double-checking. 
\item
\label{step:7}
\textbf{Recall Estimation: } Estimate the total number of positive data (output $|R_E|$ as an estimation of $|R|$).
\bi
\item
\textbf{STOP} the process and output $L_R$ if $|L_R| \ge T_{rec}|R_E|$ (where $T_{rec}$ represents the user-defined target recall).
\item
Otherwise proceed to \ref{step:8}.
\textit{This step aims at providing an accurate estimation of the current recall, so that the user could choose to stop the process when target recall $T_{rec}$ is reached.}
\ei
\item
\label{step:8}
\textbf{Query Strategy: }Apply the classifier trained in \ref{step:4} to predict on unlabeled set $(E \setminus L)$ and select $N$ data points as the queue $Q$, then go to \ref{step:3} to acquire human oracles.
\textit{This step aims at achieving highest efficiency (higher recall and lower cost) by utilizing the classifier trained in \ref{step:4} to suggest which should be reviewed/inspected by human next.}
\end{enumerate}
In the above, $N_1, N_2$ are engineering choices. $N_1$ is the batch size of the process, larger batch size usually leads to fewer times of training but also worse overall performance. $N_2 = \alpha N_1$ where $\alpha \in [0,1]$ reflects what percentage of the labeled data are double-checked. Larger $N_2$ means more double checks, which leads to more human errors covered but also higher cost. 

The active learning-based framework differs from the traditional semi-supervised/supervised learning methods by (1) starting training very early (with at least 1 positive), (2) always utilizing the trained model for sampling the data to inspect, and (3) refining the model with new inspection results. In this way, the active learning-based framework avoids wasting inspection efforts on random sampling and utilize the new inspection results to improve the model and make better predictions.

\subsection{{\IT}}

This paper is based on the above general active learning framework, and extends it for
vulnerability prediction. When applied to vulnerability prediction problem, {\IT} follows the active learning-based framework shown in Figure~\ref{fig:{\IT}} and the detailed techniques applied in each step will be presented in this subsection.

\subsubsection{Feature Extraction}
\label{sec:feature_extraction}

\textbf{Features} are not restricted in {\IT}. Any types of features extracted from the source code can be used to predict vulnerabilities in {\IT}. That said, users could extract their own features, but here we present three example types of features which are popular in existing vulnerability researches.
\bi
\item
Software metrics: followed by the work of Shin {\it et al.}~\cite{shin2011evaluating}, SciTools' Understand\footnote{http://www.scitools.com} is used to measure the static features of software.
\item
Crash features: followed by the work of Theisen {\it et al.}~\cite{Theisen2015ICSE}, the crash features measure the number of time each source code file has crashed. Such information is extracted from the crash dump stack trace data.
\item
Text mining features: different from the work of Scandariato {\it et al.}~\cite{scandariato2014predicting}, where labels (vulnerable or non-vulnerable) are required to select tokens, we apply the same text mining feature extraction as the total recall approaches~\cite{Yu2018,Yu2019}. Specifically, we:
\begin{enumerate}
\item
Tokenized all source code files without stop/control words removal or stemming.
\item
Selected tokens with largest tf-idf score across all source code files. 
For token $t$, its tf-idf score: $$\begin{aligned}\mathit{Tfidf}(t) &= \sum_{d\in D} \mathit{Tfidf}(t,d),\end{aligned}$$  
in which for token $t$ in document $d$, \[\mathit{Tfidf}(t, d)=w^t_d\times (\log \frac{|D|}{\sum_{d\in D} \mathit{sgn}(w^t_d)}+1),\] where $w^t_i$ is the number of times token $t$ appears in document $d$.
\item
Built a term frequency matrix with $M$ tokens selected---according to our prior work in retrieving relevant papers~\cite{Yu2018,Yu2019}, we use $M=4000$.
\item
Normalized each row (feature vector for each file) with their L2-norm\footnote{
L2-norm for a vector $x$ is $\sqrt{x^{T}x}$, where $T$ denotes ``transpose''}.
\end{enumerate}
\ei

\subsubsection{Initial Sampling}

Two different sampling strategies are applied in the initial sampling step of {\IT}:
\bi
\item
\textbf{Random sampling}: random sample without replacement until the first ``vulnerable'' file is found.
\item
\textbf{Domain knowledge}: apply domain knowledge to guide the initial sampling so that the first ``vulnerable'' file can be found earlier. As Theisen {\it et al.}~\cite{Theisen2015ICSE} suggested, files that have crashed before are more likely to contain security vulnerabilities. Therefore {\IT} selects files by descending order of their crash feature counts when crash data is available.
\ei

\subsubsection{Human Oracle}
\label{sec:human_oracle}
Human inspectors (software engineers) are employed to inspect the selected source code files. If vulnerabilities are found, the source code file $x$ containing the vulnerabilities is labeled as positive and is added into $L_R$. 

\subsubsection{Train Model}
\label{sec:train_model}

The train model step in {\IT} includes the following three procedures:
\begin{enumerate}
\item
Generate {\bf presumptive non-relevant examples} to alleviate the sampling bias in ``non-vulnerable'' examples ($L\setminus L_R$). 
\item
Apply \textbf{aggressive undersampling} to balance the training data. 
\item
Train a soft-margin linear \textbf{support vector machine} (SVM) on the current labeled files $L_R$ and $L$, using features extracted in \ref{step:1} to predict whether one file is vulnerable or not.
\end{enumerate}

\textbf{Presumptive non-relevant examples}  is a technique created by Cormack and Grossman~\cite{cormack2015autonomy} to alleviate the sampling bias of negative (non-vulnerable) training data. With certainty sampling, only the data close to the positive labeled data in feature space are selected for inspection. As a result, the labeled negative data are mostly close to the positive data, not spreading all over the feature space. This sampling bias of negative training data will deteriorate the performance of the trained model~\cite{cormack2015autonomy}. To alleviate such sampling bias, each time before training, {\em presumptive non-relevant examples} samples randomly from the unlabeled data ($P \subset E \setminus L$) and presumes the sampled data to be negative (non-vulnerable) in training. The rationale behind this technique is that given the low prevalence of positive (vulnerable) data, it is likely that most of the presumed ones are negative (non-vulnerable).

\textbf{Aggressive undersampling} is a data balancing technique first created by Wallace {\it et al.}~\cite{wallace2010semi}. During the training process, aggressive undersampling throws away majority (non-vulnerable) training data ($P\cup L \setminus L_R$) closest to SVM decision plane until reaching the same number of minority (vulnerable) training data $|L_R|$. Aggressive undersampling is applied only when $|L_R|\ge N_3$~\cite{Yu2018}. Here, $N_3$ is also an engineering choice to avoid aggressive undersampling reducing the number of training data too much. Later in \tion{query_strategy} we will show that $N_3$ is also the threshold where the query strategy is switched from uncertainty sampling to certainty sampling. Therefore $N_3$ also features the trade-off between faster building a better model and greedily applying the model to save inspection effort. 

\textbf{Support vector machines (SVMs)} are a well-known and widely-used classification technique. The idea behind SVMs is to map input data to a high-dimension feature space and then construct a linear decision plane in that feature space~\cite{cortes1995support}. Soft-margin linear SVM~\cite{joachims2006training} is selected as the classifier because it has been proved to be a useful model in SE text mining~\cite{krishna2016bigse} and is applied in the state-of-the-art total recall methods~\cite{Yu2018,miwa2014reducing,wallace2010semi,cormack2014evaluation} as well as the state-of-the-art vulnerability prediction methods~\cite{neuhaus2007predicting,zimmermann2010searching,nguyen2010predicting}. The soft-margin linear SVM is implemented with the SVC package from scikit-learn\footnote{https://scikit-learn.org/stable/modules/generated/sklearn.svm.SVC.html}. Default parameters are applied while kernel is set to be ``linear'' and penalty is set to be ``balanced'' (balancing the penalty of misclassifying a training data point by the population of its class, i.e. penalize more if misclassifying a minority class training data point) as suggested by Miwa {\it et al.}~\cite{miwa2014reducing}.

\textbf{Quasi-Newton Semi-Supervised Support Vector Machine (QN S3VM)}~\cite{gieseke2012sparse,gieseke2014fast} is a semi-supervised algorithm. It makes perfect sense to apply a semi-supervised learning in \ref{step:4} since there are a large number of unlabeled data available that are not utilized by supervised learners like linear SVM. S3VM is a semi-supervised version of the linear SVM, which makes it a fair comparison against the linear SVM. In addition, S3VM trains faster than other semi-supervised learners, e.g. label propagation as Meng {\it et al.}~\cite{7853039} suggested. Given that the inspector needs to wait for the model to be trained every $N_1$ files inspected, a faster learner like S3VM is preferred. 
In our experiments in \tion{exp_efficiency}, we test the S3VM learner implemented with the package\footnote{http://www.fabiangieseke.de/index.php/code/qns3vm} from Gieseke {\it et al.}~\cite{gieseke2012sparse,gieseke2014fast}.

\subsubsection{Error Prediction}
\label{sec:error_prediction}

In our prior work~\cite{Yu2019}, an error correction mechanism was designed for correcting both false negatives and false positives.  Its core assumption is:
\begin{quote}
{\em  Human classification errors are more likely to be found where human and machine classifiers disagree most.}
\end{quote}
Since the error distribution of the vulnerability inspection process is different from that of selecting research papers (only false negatives are considered), it cannot be directly applied to vulnerability prediction. As a result, we adopt the same assumption and designed \textbf{DISPUTE}, which focuses on correcting false negatives only.

\textbf{DISPUTE} applies the model trained in \ref{step:4} to predict on the source code files that have only been inspected once and are labeled as ``non-vulnerable'' ($L \setminus L_R$). Select $N_2$ files with highest prediction probability of being vulnerable and add the selected files into a queue for double-checking.

\subsubsection{Double check}

Recall from the introduction that
when humans   inspect a file, they may  fail to find vulnerabilities~\cite{hatton2008testing}. 
Some double checks (where files are  examined by more than one human)
are required to cover such errors. 
Later in this paper(\tion{errors}), we will test two strategies for double-checking:
\bi
\item
\textbf{DISPUTE} double-checks each file selected in \ref{step:5} (of \fig{{\IT}}) once.
\item
\textbf{DISPUTE(3)} double-checks each file selected in \ref{step:5} twice if the first double-checker still finds it to be ``non-vulnerable''.
\ei
\textbf{DISPUTE(3)} reduces the chance of the selected file being mislabeled again at the cost of doubling the double-checking effort. This strategy is especially useful when the human inspectors have high chance to miss vulnerable files.

\begin{algorithm}[!htp]
\scriptsize
\SetKwInOut{Input}{Input}
\SetKwInOut{Output}{Output}
\SetKwInOut{Parameter}{Parameter}
\SetKwRepeat{Do}{do}{while}
\Input{$E$, set of all candidates\\$L$, set of labeled data\\$L_R$, set of labeled positive data\\$CL$, classifier trained in \ref{step:4}}
\Output{$|R_E|$, estimated number of positive data}
\BlankLine

\Fn{SEMI ($CL,E,L,L_R$)}{
    $|R_E|_{last}\leftarrow 0$\;
    $\neg L \leftarrow E \setminus L$\;
    \BlankLine
    \ForEach{$x \in E$}{
        $D(x) \leftarrow CL.decision\_function(x)$\;
        \If{$x \in |L_R|$}{
            $Y(x)\leftarrow 1 $\;
        }
        \Else{
            $Y(x)\leftarrow 0 $\;
        }
    }
    \BlankLine
    $|R_E| \leftarrow \sum\limits_{x\in E} Y(x)$\;
    \BlankLine
    \While{$|R_E|\neq |R_E|_{last}$}{
    \BlankLine
       \tcp{Fit and transform Logistic Regression}
       $LReg \leftarrow LogisticRegression(D,Y)$\;
       \BlankLine
       $Y \leftarrow TemporaryLabel(LReg,\neg L,Y)$\;
       \BlankLine
       $|R_E|_{last}\leftarrow |R_E|$\;
       \BlankLine
       \tcp{Estimation based on temporary labels}
       $|R_E| \leftarrow \sum\limits_{x\in E} Y(x)$\;
    }
    \Return{$|R_E|$}\;
}
\BlankLine
\Fn{TemporaryLabel ($LReg,\neg L,Y$)}{
    $count \leftarrow 0$\;
    $target \leftarrow 1$\;
    $can \leftarrow \emptyset$\;
    \BlankLine
    \tcp{Sort $\neg L$ by descending order of $LReg(x)$}
    $\neg L \leftarrow SortBy(\neg L,LReg)$\;
    \BlankLine
    \ForEach{$x \in \neg L$}{
        $count \leftarrow count+LReg(x)$\;
        $can \leftarrow can \cup \{x\}$\;
        \If{$count \geq target$}{
            $Y(can[0]) \leftarrow 1$\;
            $target \leftarrow target+1$\;
            $can \leftarrow \emptyset$\;
        }
    }
    \Return{$Y$}\;
}

\caption{Psuedo Code for SEMI~\cite{Yu2019}}\label{alg:SEMI}
\end{algorithm}

\subsubsection{Recall Estimation}
\label{sec:recall_estimation}

In our prior work~\cite{Yu2019}, a semi-supervised estimator called \textbf{SEMI} is designed for estimating current recall. SEMI utilizes a recursive \emph{TemporaryLabel} technique. Each time the \textbf{SVM} model is retrained, SEMI assigns temporary labels to unlabeled examples and builds a logistic regression model on the temporary labeled data. It then uses the obtained regression model to predict the likelihood of being positive of the unlabeled data and updates the temporary labels. This process is iterated until convergence and the final temporary labels are used as an estimation of the total number of positive examples in the dataset. Detailed algorithm of SEMI is shown in Algorithm~\ref{alg:SEMI}.

\subsubsection{Query Strategy}
\label{sec:query_strategy}

{\em Query strategy is a key part of active learning.} The two query strategies applied in {\IT} are: 1) {\bf uncertainty sampling}, which picks the unlabeled examples that the active learner is most uncertain about (examples that are closest to the SVM decision hyperplane); and 2) {\bf certainty sampling}, which picks the unlabeled examples that the active learner is most certain to be vulnerable (examples on the vulnerable side of and are furthest to the SVM decision hyperplane). Different researchers endorse different query strategies, e.g. Wallace {\it et al.}~\cite{wallace2010semi} applies uncertainty sampling in their patient active learning, Cormack {\it et al.}~\cite{cormack2014evaluation,miwa2014reducing} use certainty sampling from beginning to end. Followed by our previous design for literature review~\cite{Yu2019}, {\IT} applies uncertainty sampling early on and certainty sampling afterward.

More specifically, \ref{step:8}
(of \fig{{\IT}}) applies the model trained in \ref{step:4} to predict on unlabeled files $(E \setminus L)$ and 
\bi
\item
select top $N_1$ files with vulnerability prediction probability closest to $0.5$ ({\bf uncertainty sampling}) if $|L_R|< N_3$; or
\item
select top $N_1$ files with vulnerability prediction probability closest to $1.0$ ({\bf certainty sampling}) if $|L_R|\ge N_3$. 
\ei
Push the selected files into the queue $Q$, then go to \ref{step:3} for another round of inspection.

\begin{figure*}[!p]
    \centering
    \subfloat[\textbf{Step 1}: initialize each data points with 2 dimensionality features extracted from the source code of each file.]
    {
        \includegraphics[width=0.22\linewidth]{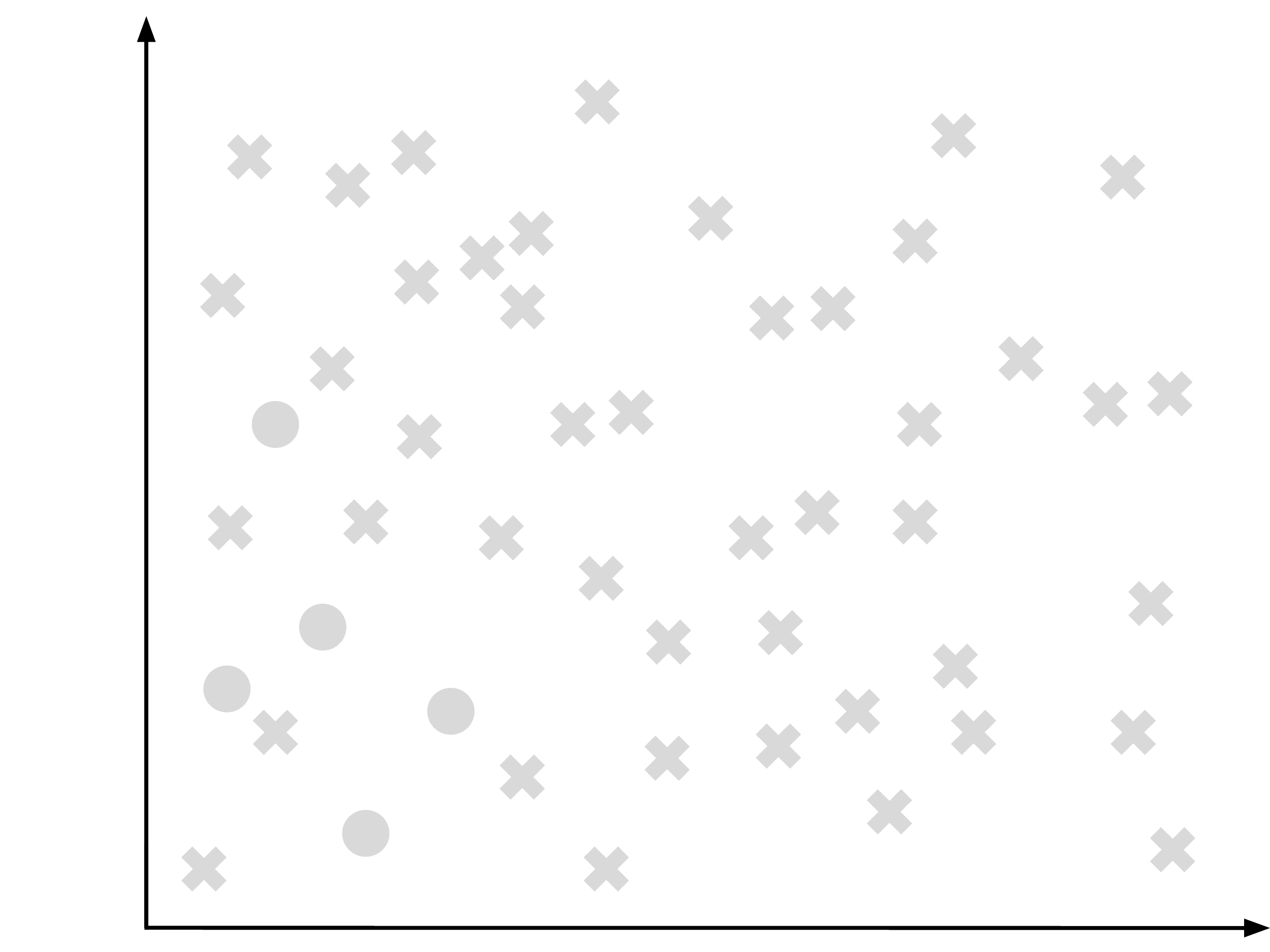}
    }\quad
    \subfloat[\textbf{Step 2}: random sample $N_1=2$ data points (in hollow green) for human oracles.]
    {
        \includegraphics[width=0.22\linewidth]{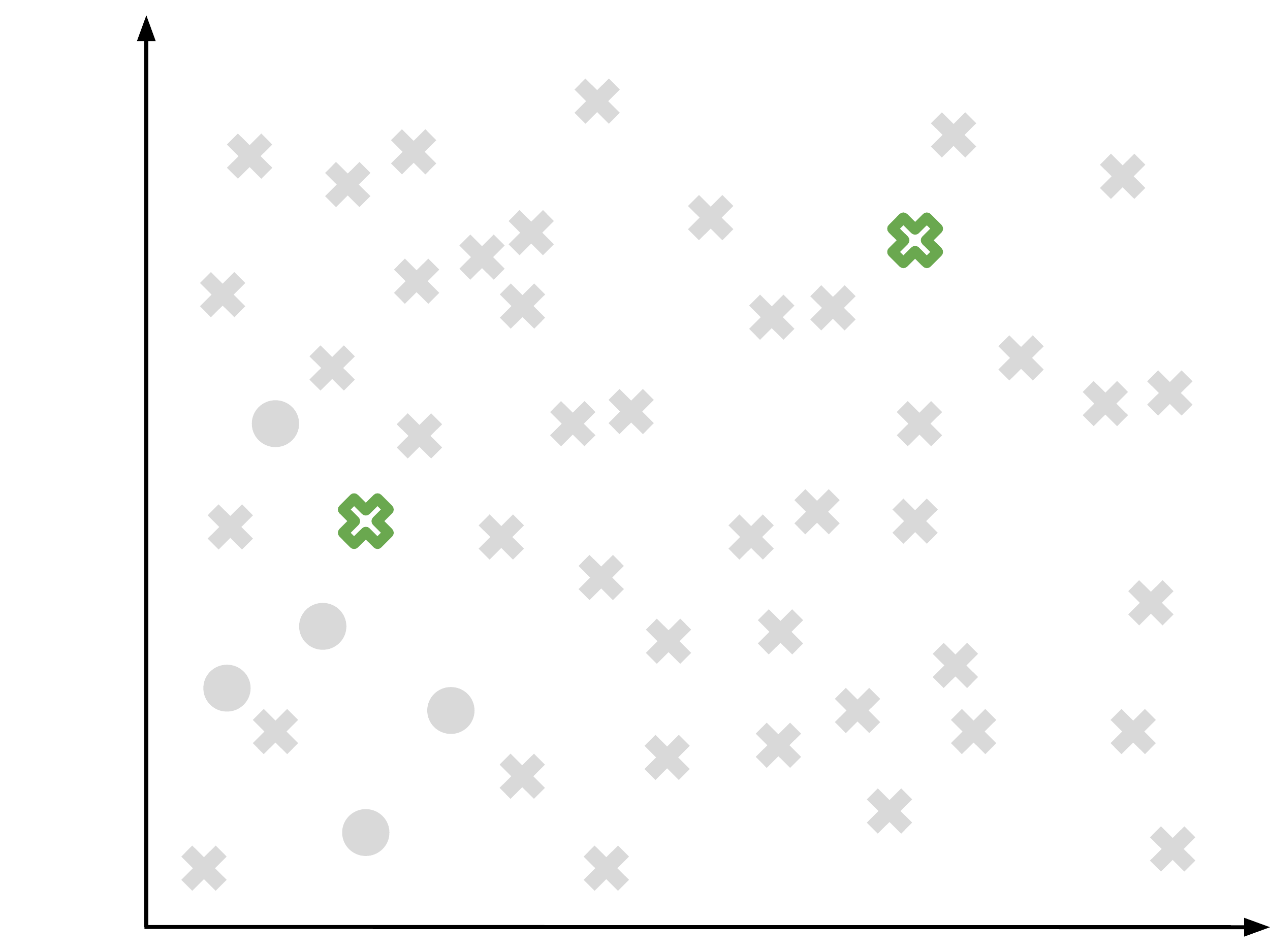}
    }\quad
    \subfloat[\textbf{Step 3}: human inspects the selected $N_1=2$ points and labels them as negative (blue). Since $|L_R|<1$, return to \textbf{Step 2}.]
    {
        \includegraphics[width=0.22\linewidth]{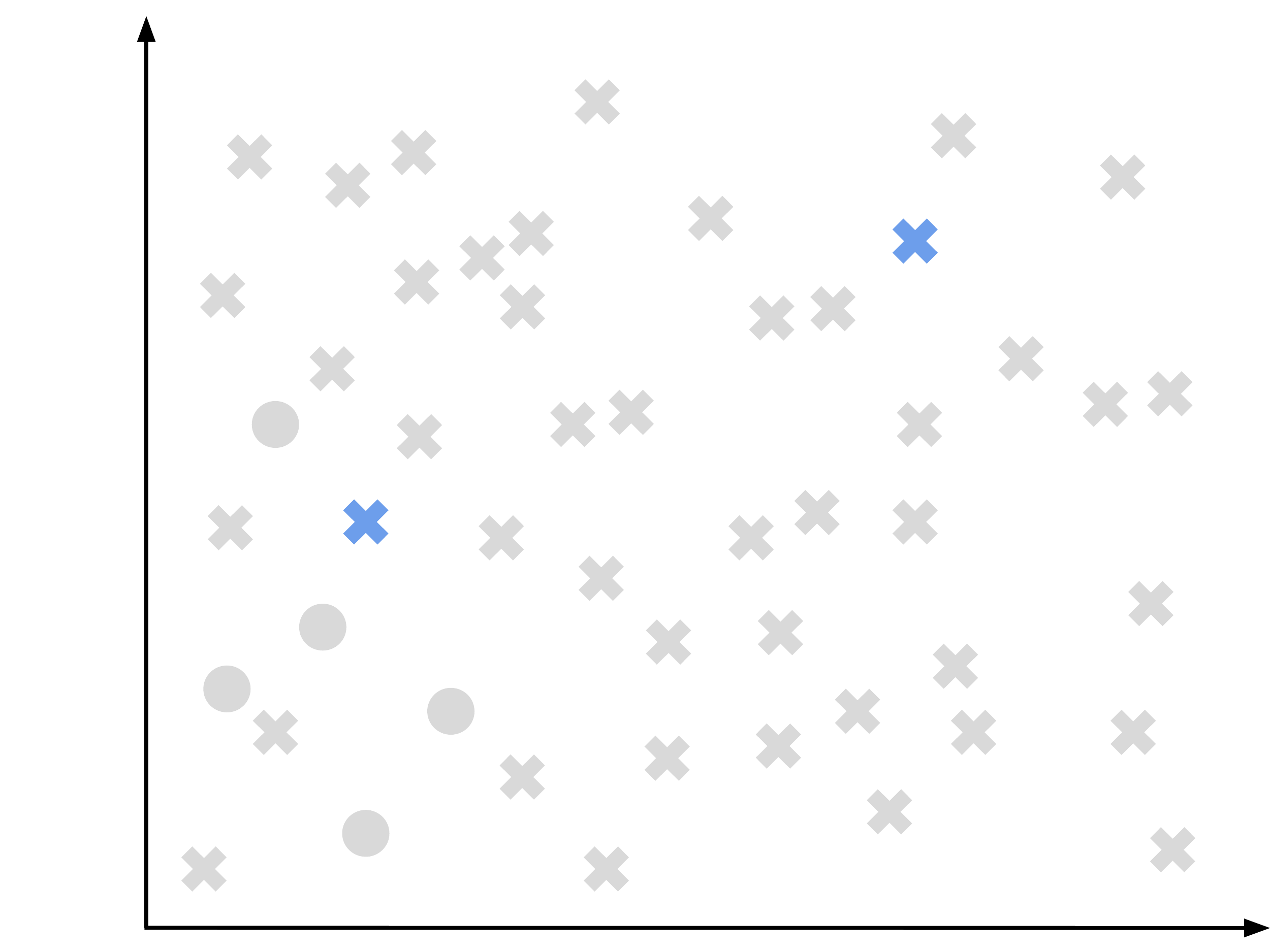}
    }\quad
    \subfloat[\textbf{Step 2}: random sample $N_1=2$ data points (in hollow green) for human oracles.]
    {
        \includegraphics[width=0.22\linewidth]{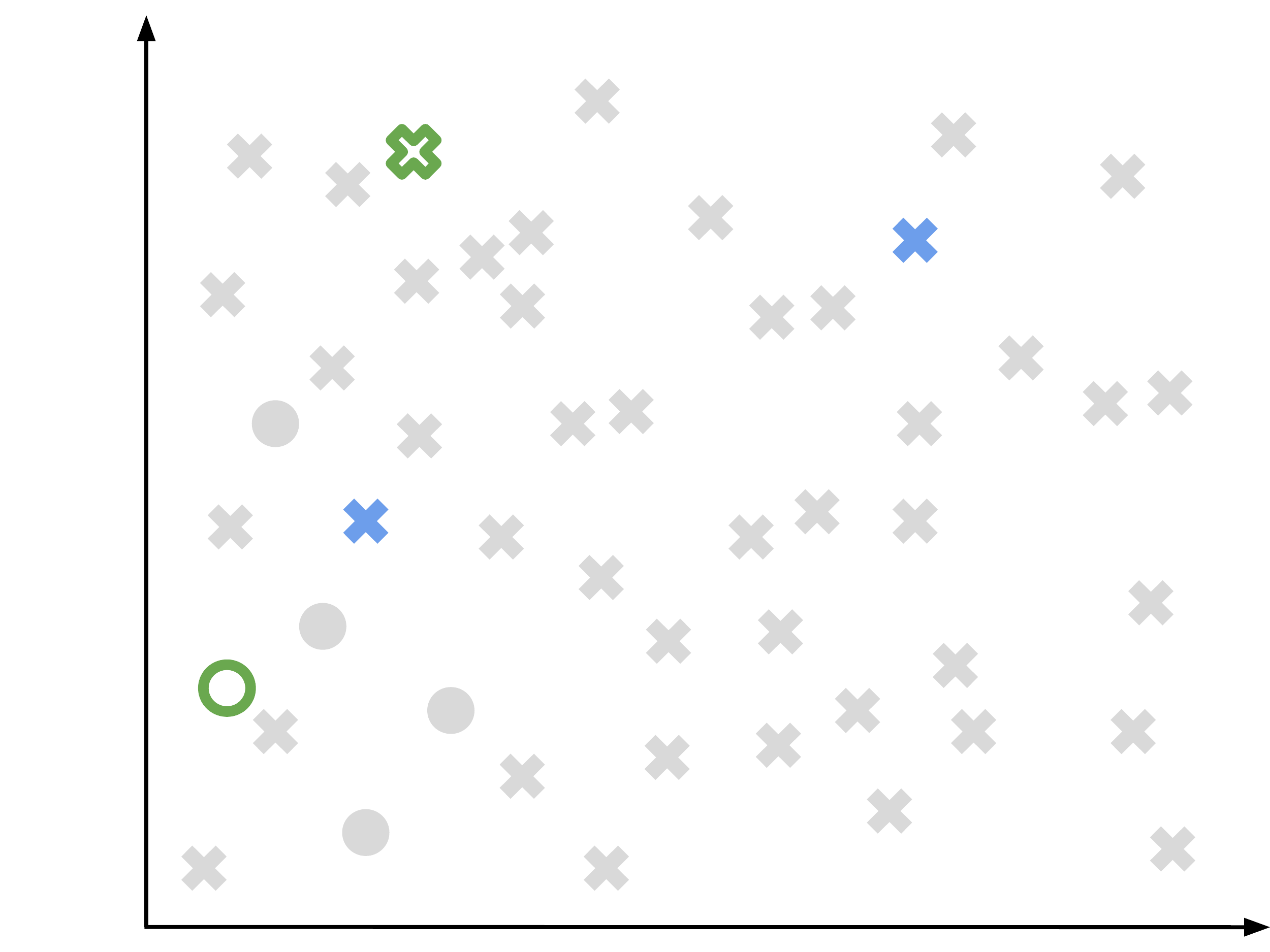}
    }\\
    \subfloat[\textbf{Step 3}: human inspects the selected $N_1=2$ points and labels them as positive (red) and negative (blue). Since $|L_R|\ge 1$, proceed to \textbf{Step 4}.]
    {
        \includegraphics[width=0.22\linewidth]{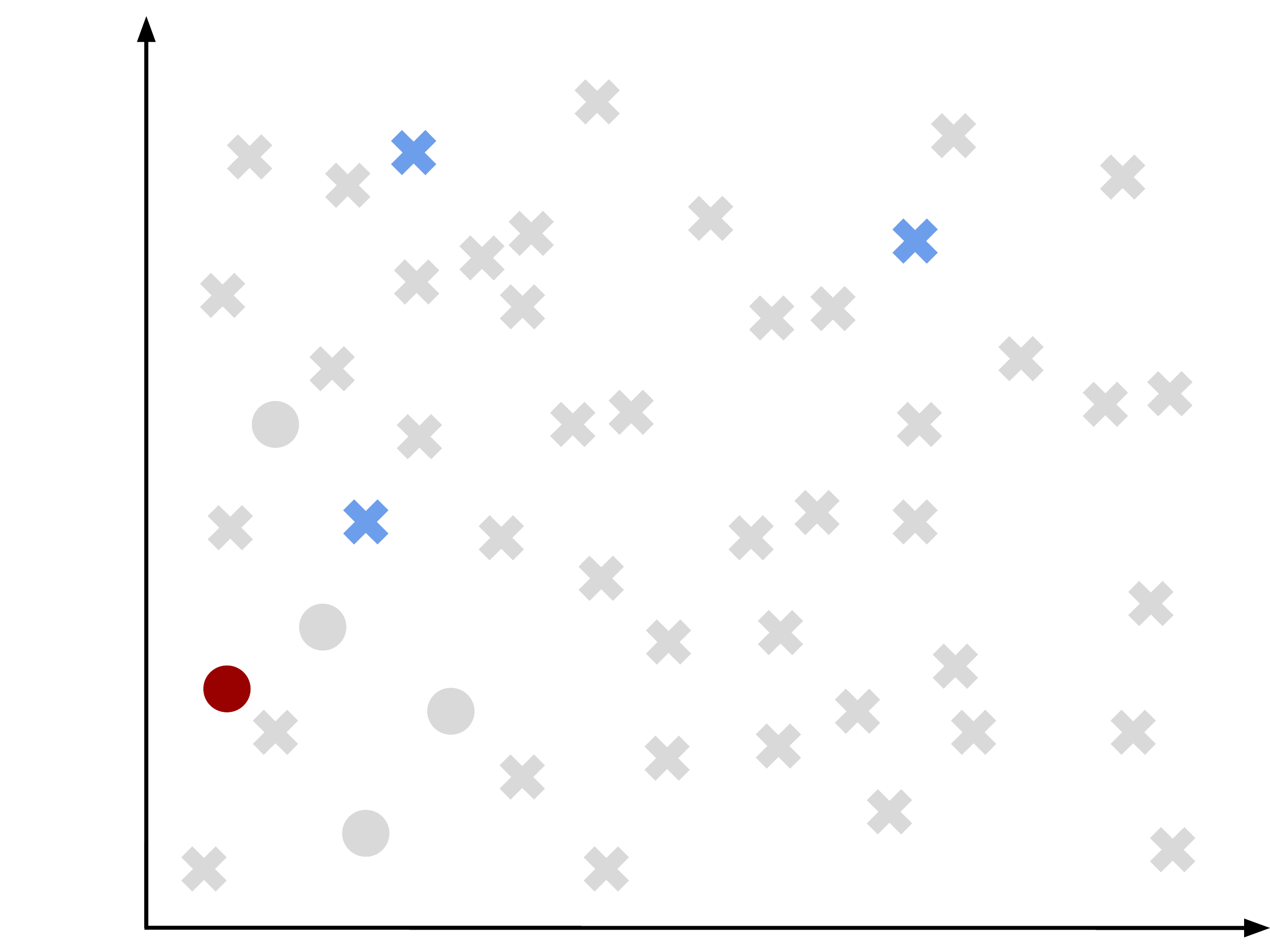}
    }\quad
    \subfloat[\textbf{Step 4}: since $|L_R|<N_3=2$, train a linear SVM model with balanced weights and presumptive negative examples (hollow blue \textbf{X}s). The SVM decision plane is shown as the black line.]
    {
        \includegraphics[width=0.22\linewidth]{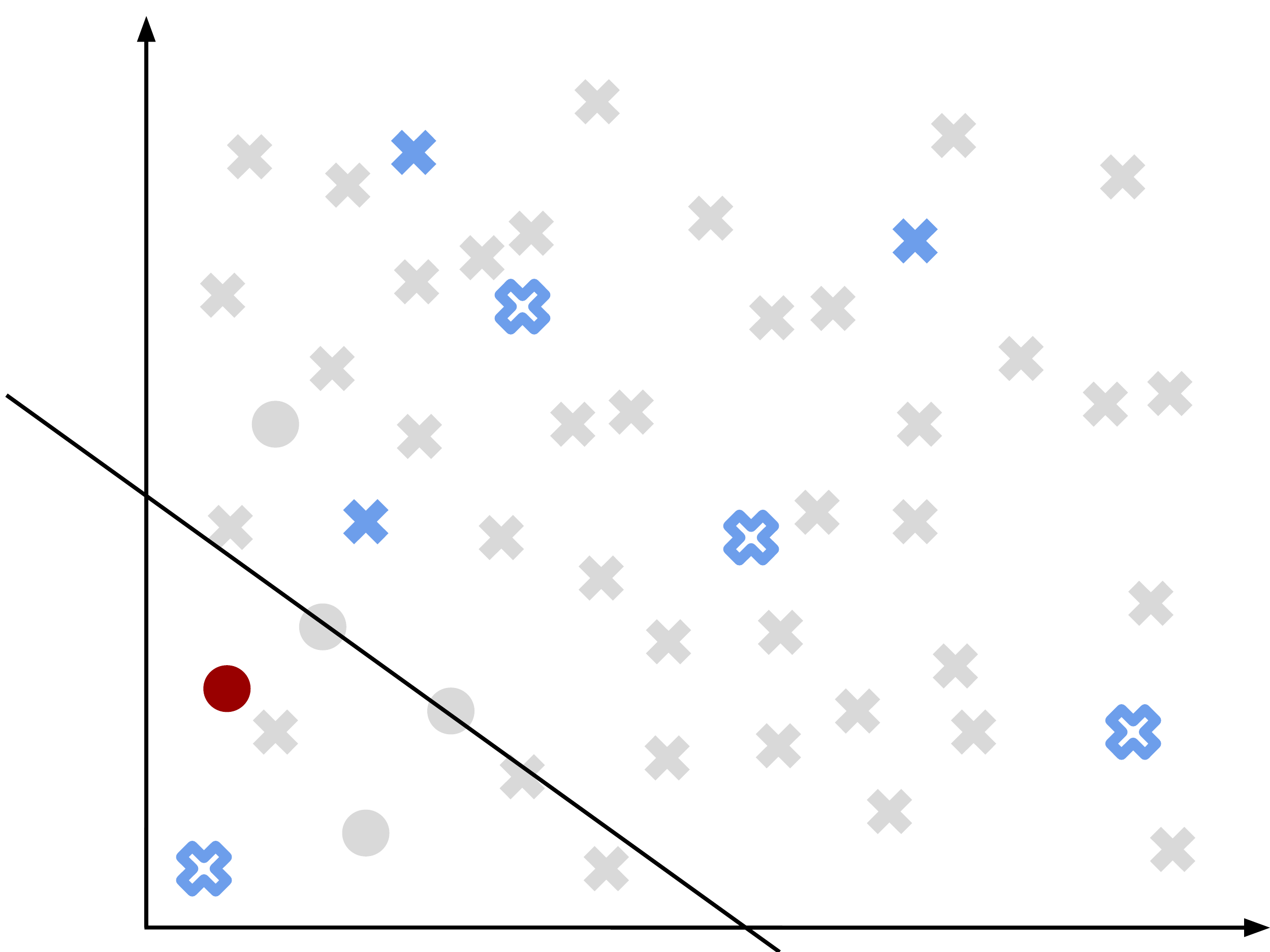}
    }\quad
    \subfloat[\textbf{Step 5}: apply the SVM model for error prediction, select $N_2=1$ data point (in hollow green) with highest prediction probability to be positive for double-checking.]
    {
        \includegraphics[width=0.22\linewidth]{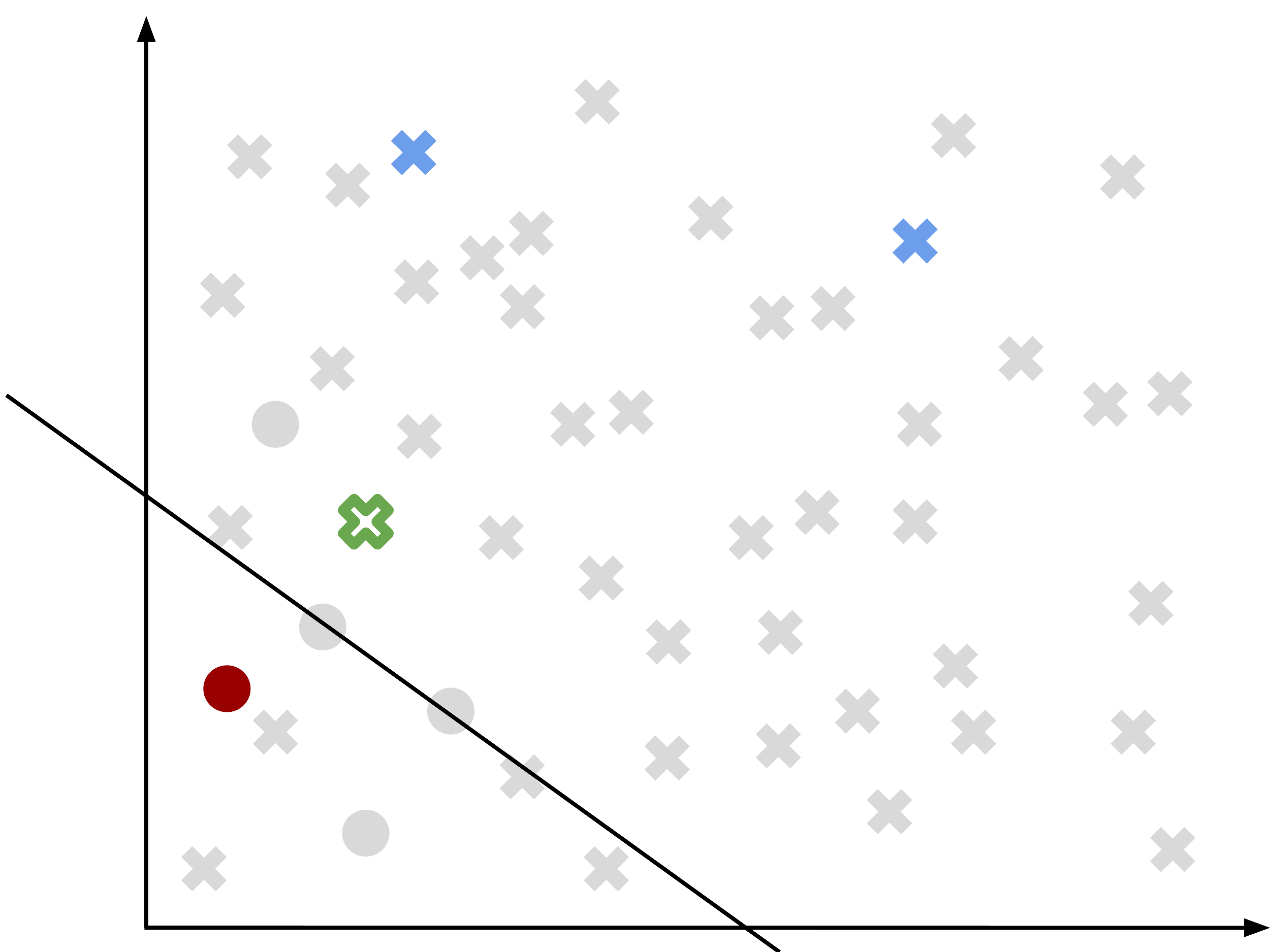}
    }\quad
    \subfloat[\textbf{Step 6}: human double-checks the selected $N_2=1$ data point and labels it as negative (blue). \textbf{Step 7}: when converges, SEMI temporarily labels 5 data points as positive (hollow red), so $|R_E|=5+|L_R|=6$. Since $|L_R|<T_{rec}|R_E|=4.8$, proceed to \textbf{Step 8}.]
    {
        \includegraphics[width=0.22\linewidth]{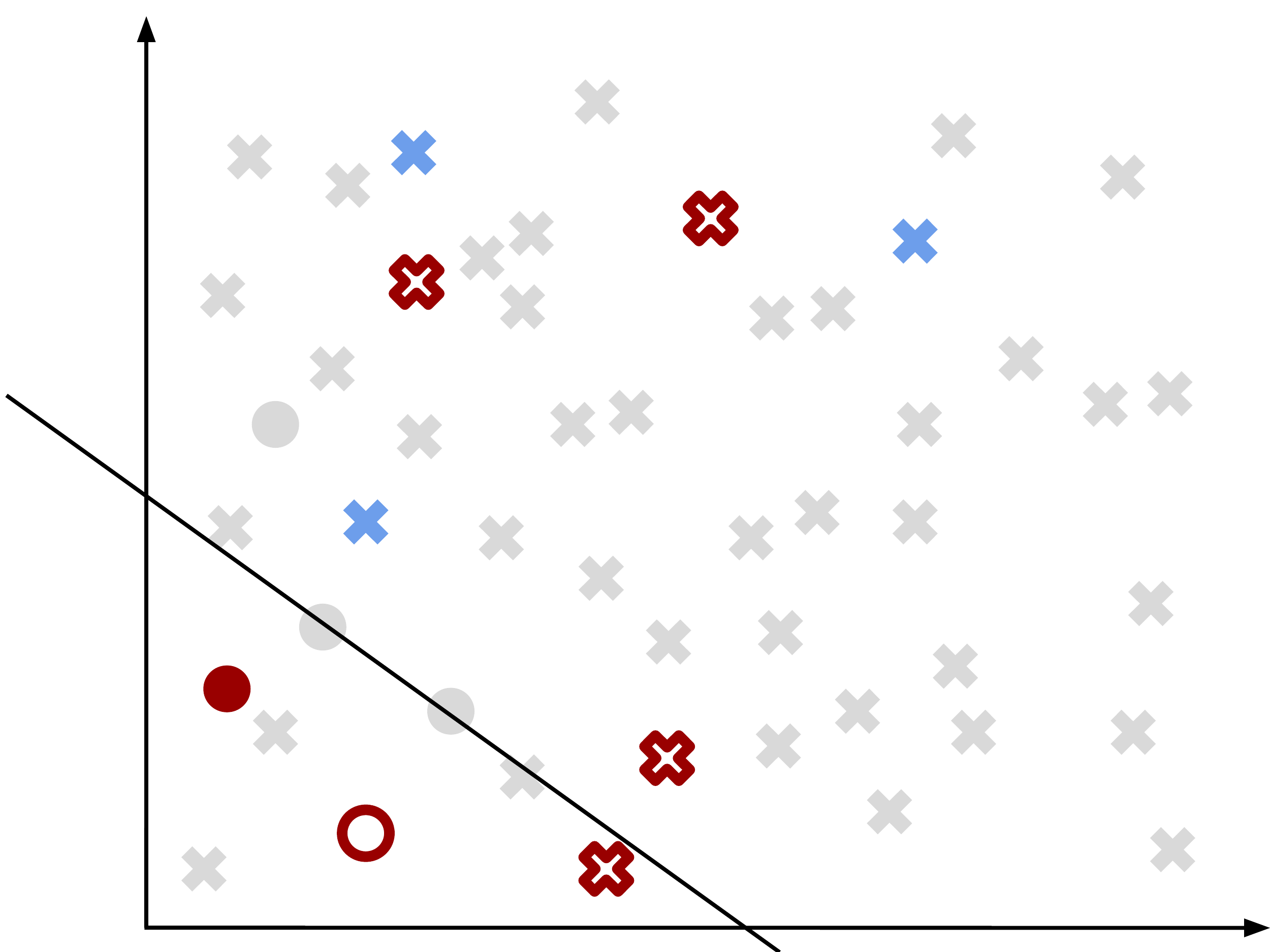}
    }\\
    \subfloat[\textbf{Step 8}: since $|L_R|<N_3=2$, apply uncertainty sampling to select $N_1=2$ points (in hollow green) closest to the SVM decision plane for human oracles.]
    {
        \includegraphics[width=0.22\linewidth]{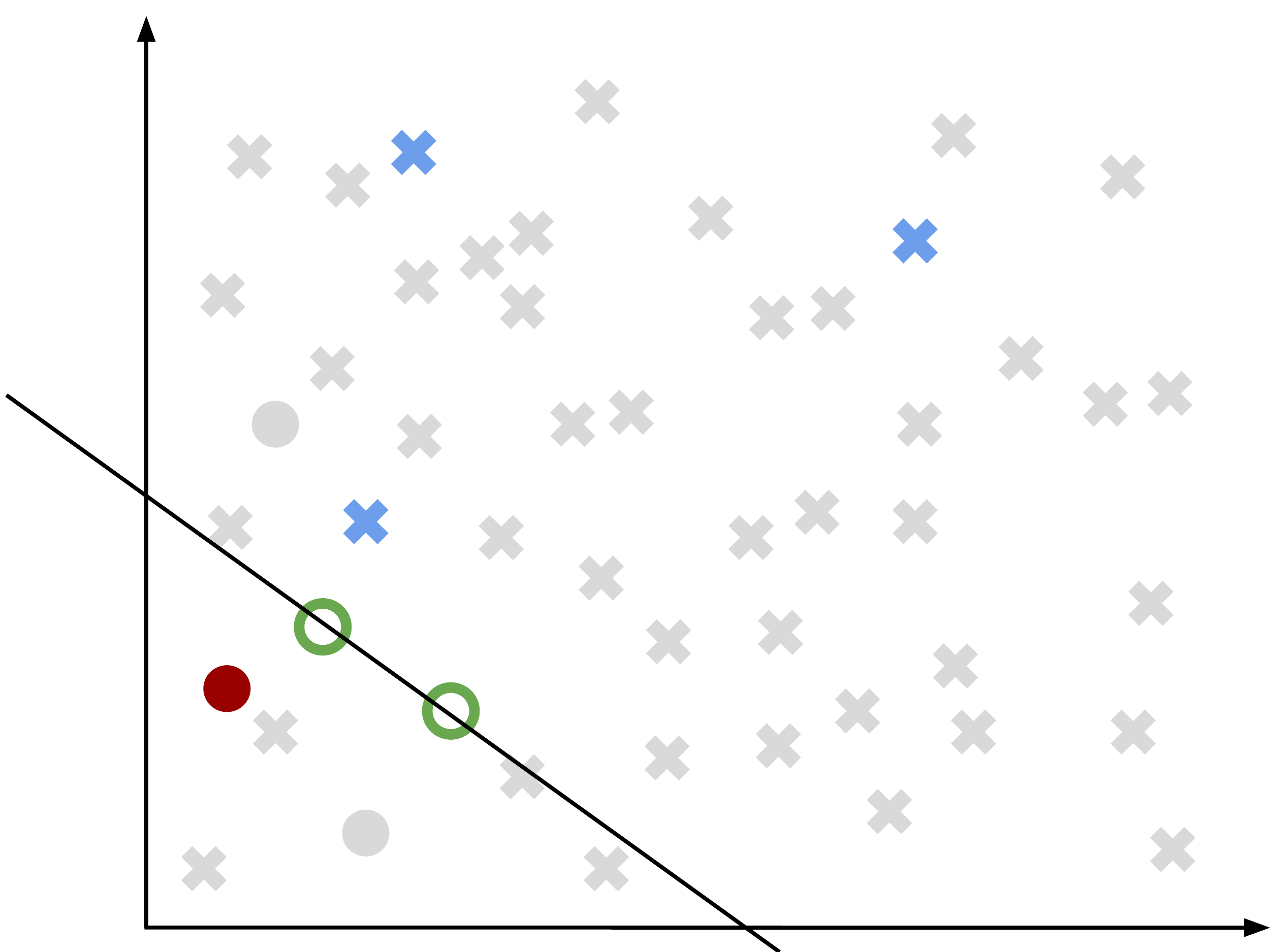}
    }\quad
    \subfloat[\textbf{Step 3}: human inspects the selected $N_1=2$ points and labels them as positive (red) and negative (blue). This time, human wrongly labels one data point.]
    {
        \includegraphics[width=0.22\linewidth]{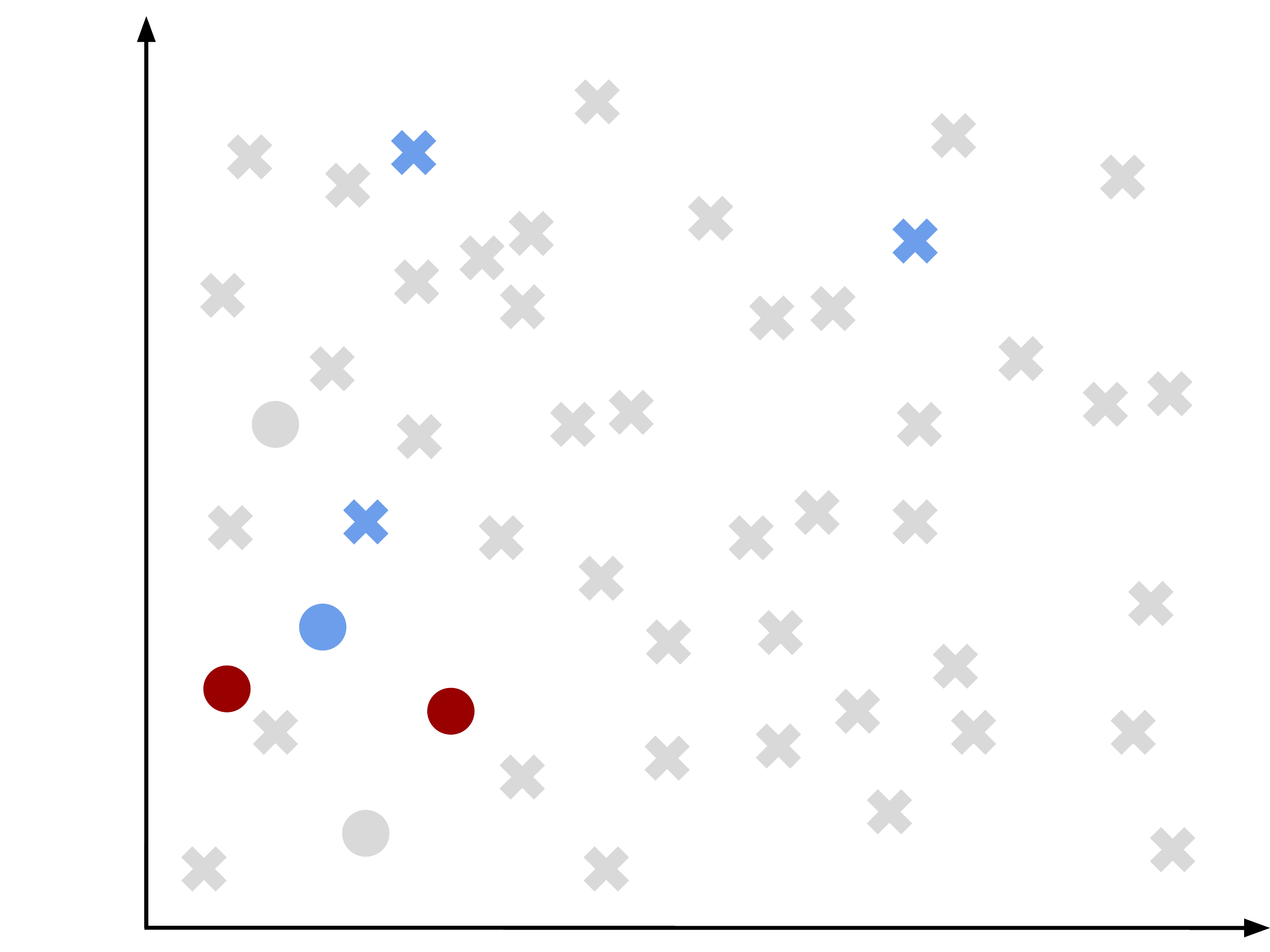}
    }\quad
    \subfloat[\textbf{Step 4}: since $|L_R|\ge N_3=2$, train a linear SVM model with presumptive negative examples (hollow blue \textbf{X}s) and aggressive undersampling (keep $|L_R|=2$ negatives).]
    {
        \includegraphics[width=0.22\linewidth]{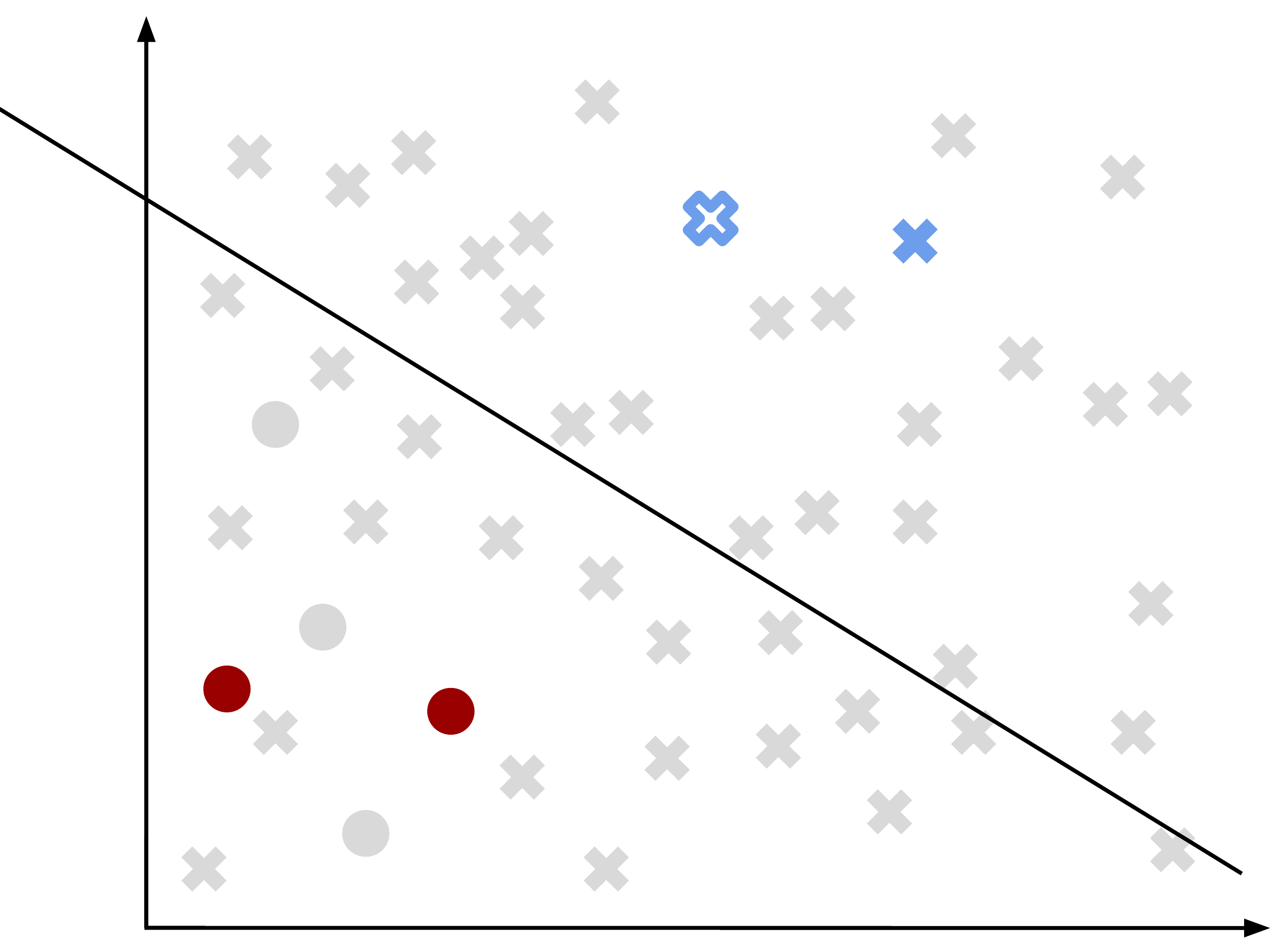}
    }\quad
    \subfloat[\textbf{Step 5}: apply the SVM model for error prediction, select $N_2=1$ data point (in hollow green) with highest prediction probability to be positive for double-checking.]
    {
        \includegraphics[width=0.22\linewidth]{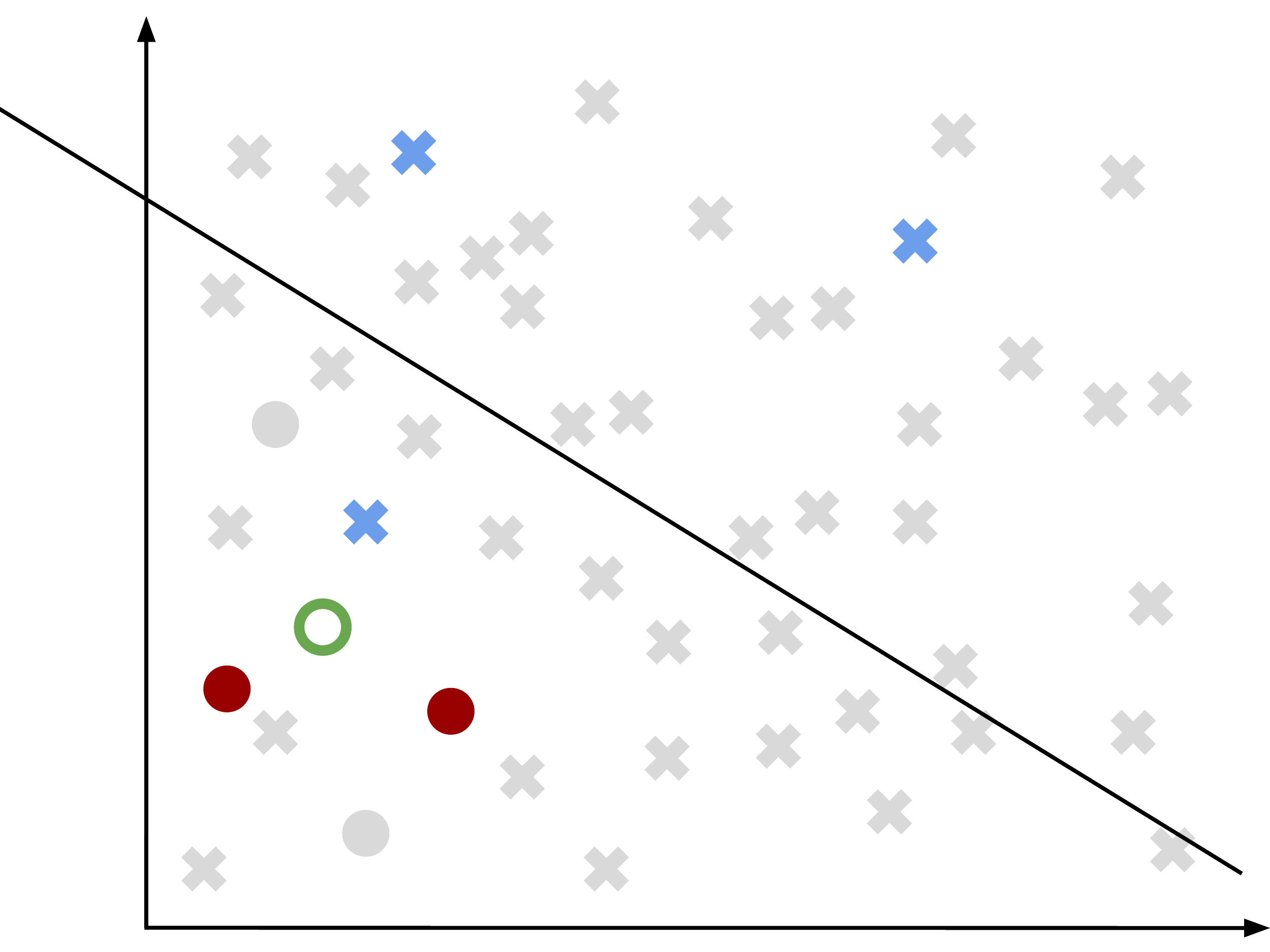}
    }\\
    \subfloat[\textbf{Step 6}: human double-checks the selected $N_2=1$ data point and label it as positive (red). This time, false negative is corrected. ]
    {
        \includegraphics[width=0.22\linewidth]{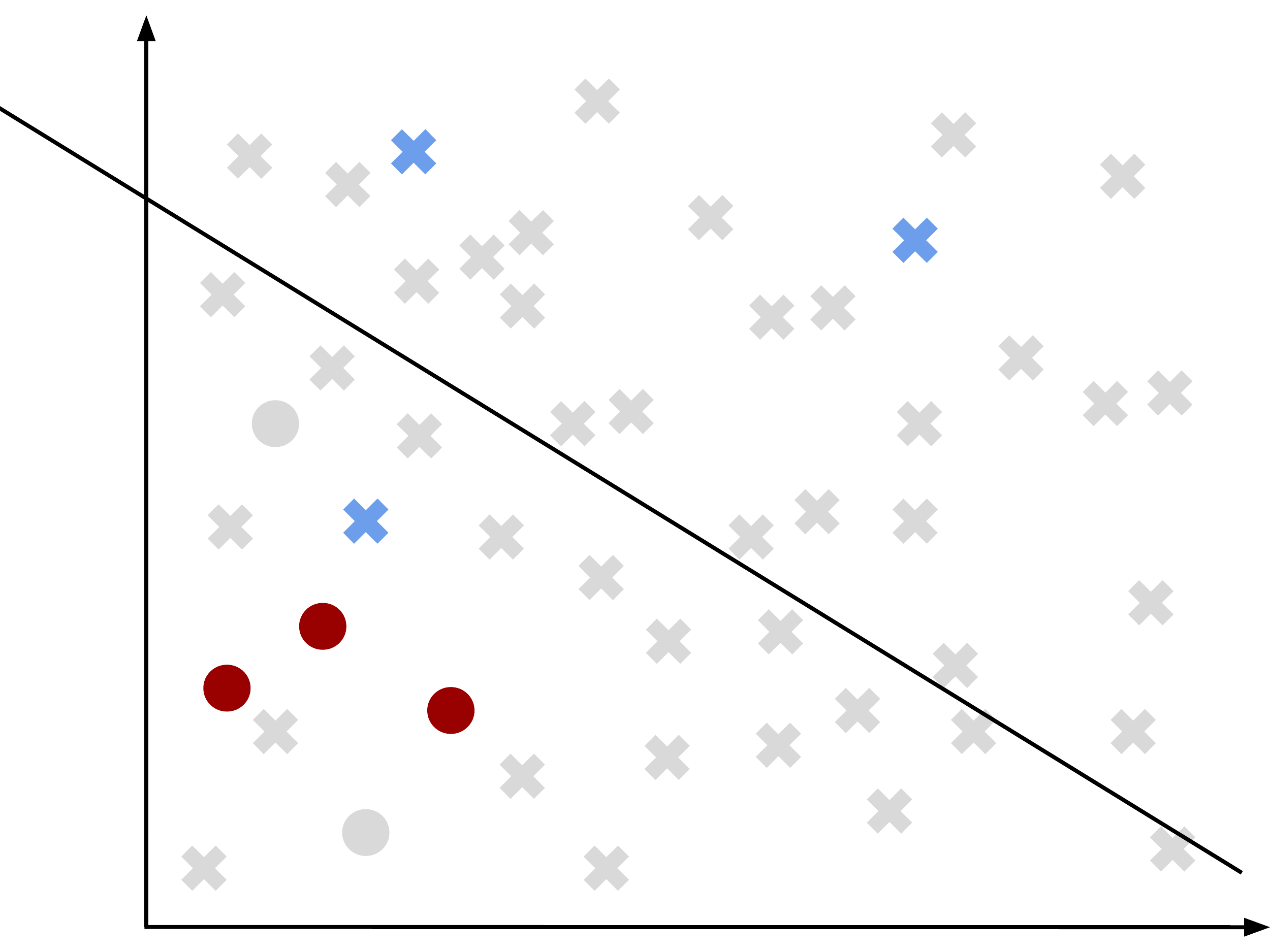}
    }\quad
    \subfloat[\textbf{Step 7}: when converges, SEMI temporarily labels 4 data points as positive (hollow red), so $|R_E|=4+|L_R|=7$. Since $|L_R|<T_{rec}|R_E|=5.6$, proceed to \textbf{Step 8}.]
    {
        \includegraphics[width=0.22\linewidth]{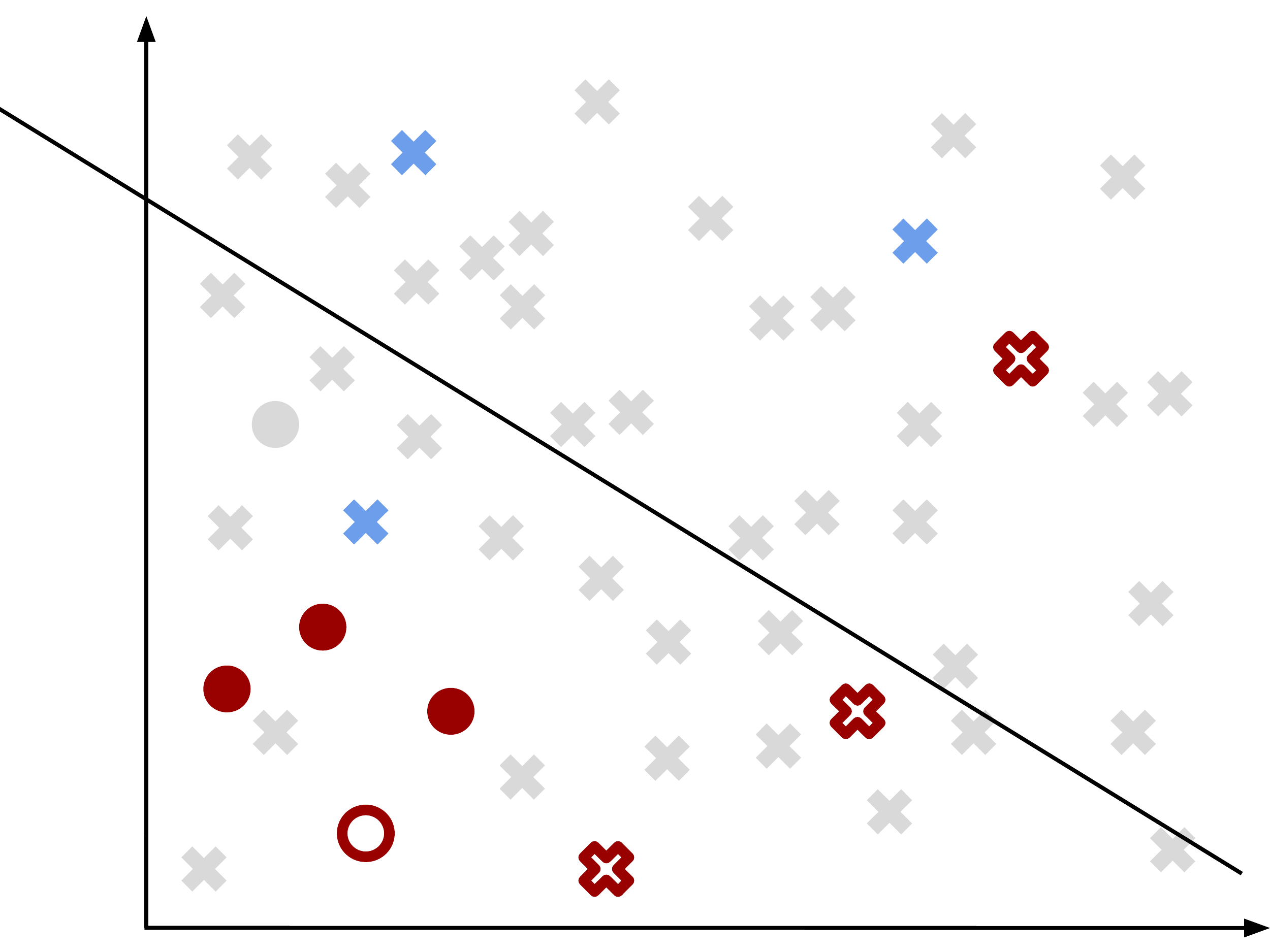}
    }\quad
    \subfloat[\textbf{Step 8}: since $|L_R|\ge N_3=2$, apply certainty sampling to select $N_1=2$ points (in hollow green) with highest prediction probability to be positive for human oracles.]
    {
        \includegraphics[width=0.22\linewidth]{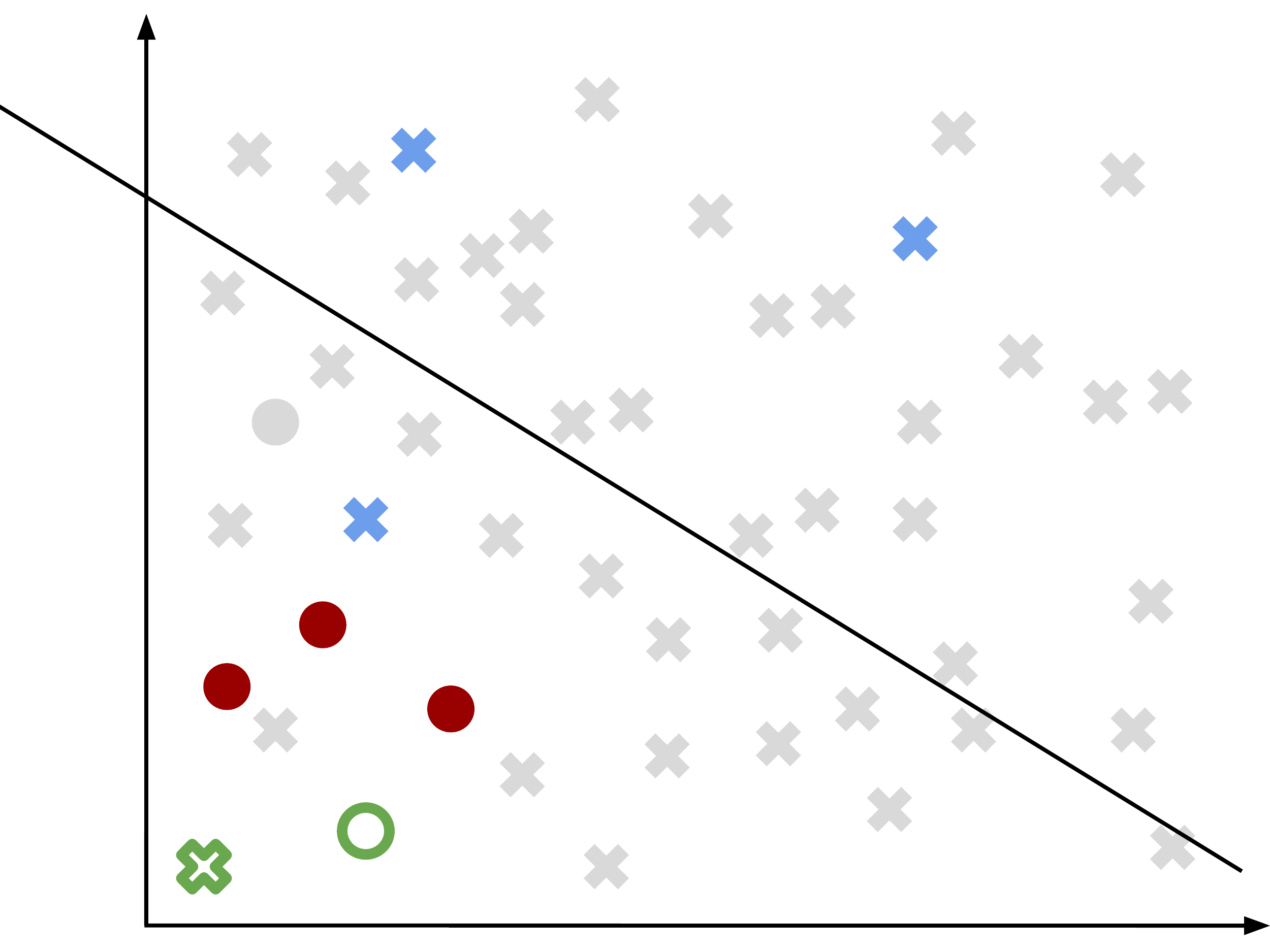}
    }\quad
    \subfloat[Iterate for 7 more rounds, until \textbf{Step 7}: when converges, SEMI temporarily labels 1 data points as positive (hollow red), so $|R_E|=1+|L_R|=5$. Since $|L_R|\ge T_{rec}|R_E|=4$, stop and output the $|L_R|=4$ points (in red). ]
    {
        \includegraphics[width=0.22\linewidth]{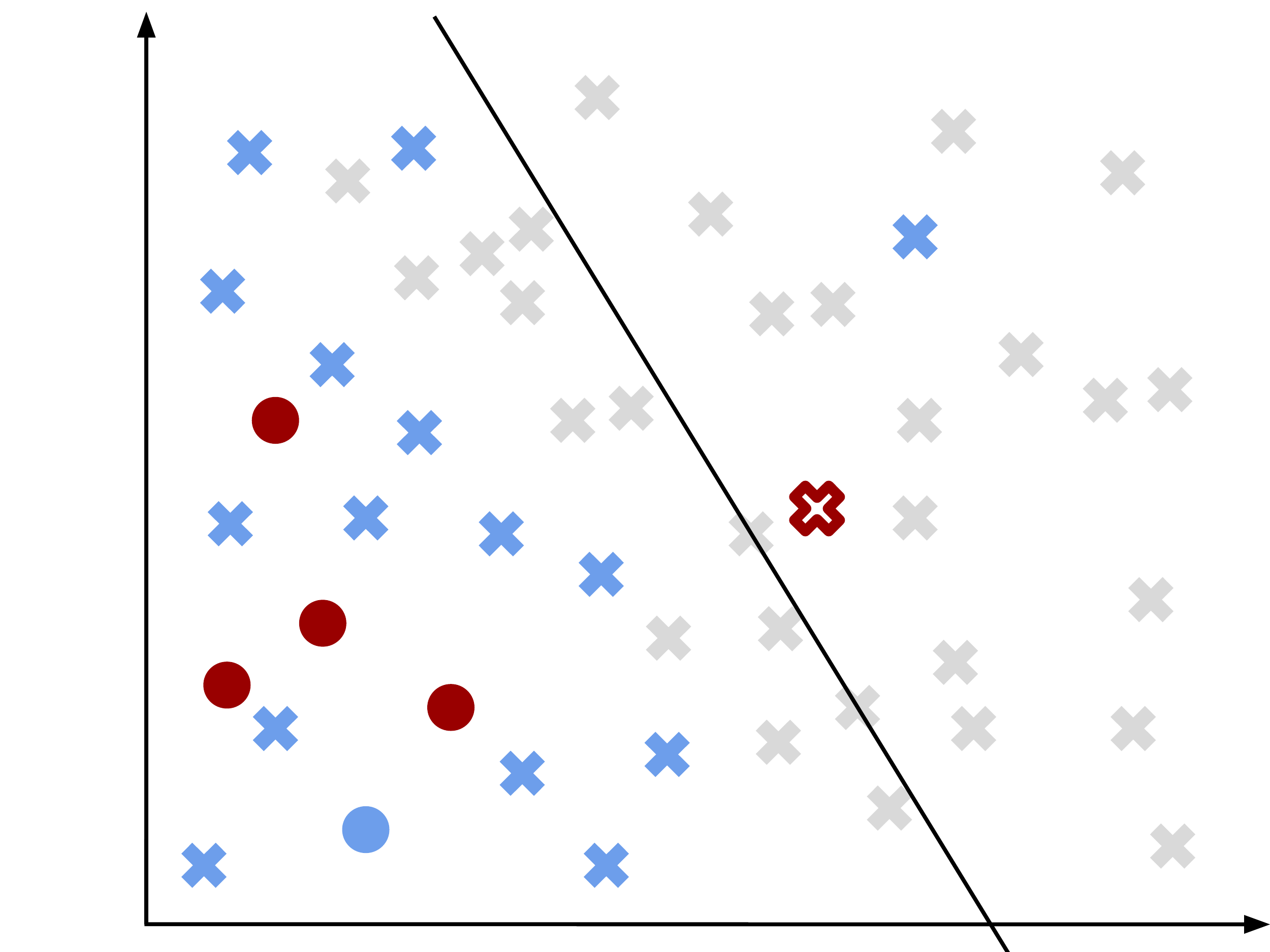}
    }
    \caption{Consider $|E| = 50$ source code files in a software project. If human inspects all the files correctly, they will find $|R|=5$ files (\textbf{O}s) being positive (vulnerable) and $|E\setminus R|=45$ files (\textbf{X}s) being negative (vulnerability-free). In this demo, {\IT} guides humans to inspect 20 (40\%) source code files and find 4 (80\%) of the vulnerable files. Data points colored in \colorbox{myGray}{\textcolor{white}{gray}} are unlabeled $E\setminus L$, while data points colored in \colorbox{myRed}{\textcolor{white}{red}} are labeled as positive $L_R$ and data points colored in \colorbox{myBlue}{\textcolor{white}{blue}} are labeled as negative $L\setminus L_R$. As for the hollow ones, \colorbox{myGreen}{\textcolor{white}{green}} are the data points selected for inspections/double checks while \colorbox{myRed}{\textcolor{white}{red}}/\colorbox{myBlue}{\textcolor{white}{blue}} are the data points temporarily labeled as positive/negative by SEMI/presumptive negative examples. Black lines are the decision hyper-planes of the linear SVM model. }
    \label{fig:demo}
\end{figure*}

\subsubsection{Summary}

A simple demonstration of the {\IT} inspection process is shown in Figure~\ref{fig:demo}. In this figure,  {\IT} incrementally updates its model and uses it to guide the human inspector to only inspect the most informative files:
\bi
\item
The demo is progressive, i.e., from top left,
 we see the start of the process. In those initial
 plots,  every item is a file. The subsequent plots show how {\IT} guides the human inspector to reveal vulnerabilities.
\item
Any item with a solid color represents a file where humans have inspected.
\item
The hollow green items in Figure~\ref{fig:demo} 
are files that {\IT} suggests the human to inspect.
For example, these hollow green items appear in (b), (g).
Subsequently,  in (c), (h) we see the labels offered by humans (see the blue and red solid items).
\item
The solid diagonal lines show updates to the decision hyper-plane of the SVM model. Note that this line first appears in (f) and is then updated in (k) and (p).
\ei
In the demo, for simplicity, parameters are set to $N_1=2,N_2=1,N_3=2$ and the feature dimensionality is set to be 2, i.e. each data point (source code file) is represented by 2 features. Target recall is set to be $T_{rec}=0.8$ and positive data points can be sometimes mislabeled as negative to demonstrate human errors.

\begin{table*}[!tbh]
\caption{Descriptive statistics for vulnerabilities types grouping}
\label{tab:data}
\small
\centering
\begin{threeparttable}
\begin{tabular}{l|c|l}
Vulnerability Type  & \makecell{\#Vulnerable Files} & Containing Types\\ \hline
\rowcolor{gray!25}Protection Mechanism Failure      & 119               & protection mechanism failure. \\
Resource Management Errors    & 85                & \makecell[l]{uncontrolled resource consumption, improper resource shutdown or release, \\  resource management errors, use after free, resource leak.}\\
\rowcolor{gray!25}Data Processing Errors         & 35                & data processing errors. \\
Code Quality        & 29                & code quality. \\
\rowcolor{gray!25}Other & 32 &  race conditions, configuration, environment, traversal, link-following, other.\\
All                 & 271               & All 14 types of vulnerabilities.\\  \hline
\end{tabular}
\begin{tablenotes}
The Mozilla Firefox dataset contains 14 different types of vulnerabilities and some files are associated with multiple types of vulnerabilities. To simulate security reviews targeting specific types of vulnerabilities, we group the vulnerabilities into five categories to ensure that each category contains enough data (\#Vulnerable Files) for a simulation. Column \#Vulnerable Files reports the number of unique files containing the specific type of vulnerabilities. Files containing multiple vulnerabilities still count for 1 but can contribute to multiple groups.
\end{tablenotes}
\end{threeparttable}
\end{table*}

\section{Mozilla Firefox Case Study}
\label{sec:case-study}

While \tion{methodology} shows how {\IT} should be applied in practice with humans inspecting codes, it is too expensive for human inspectors to test different treatments and answer all the research questions. As a result, the performance of {\IT} is tested through simulations on the Mozilla Firefox vulnerability dataset.

\subsection{Dependent Variables}
\label{sec:dependent}

The Mozilla Firefox dataset focuses on C and C++ files in Mozilla Firefox. Metrics were collected for 28,750 unique source code files in the project, within which 135 vulnerabilities affecting 271 files are manually labeled as the ground truth. These ground truth vulnerabilities were manually collected from Mozilla Foundation Security Advisories blog~\cite{mozillablog} and bug reports from Bugzilla\footnote{https://bugzilla.mozilla.org/} up  to November 21st, 2017.  As for Mozilla Foundation Security Advisories blog, vulnerability type and source code file(s) modified to fix the vulnerability were recorded. From this source, 247 files are found to contain vulnerabilities. Mozilla's bug database was then mined for bugs that were not reported publicly on the blog as vulnerabilities. Two human raters individually read through each bug report and classified each as ``vulnerability'' or ``not a vulnerability''. The two raters looked at the description of each bug, along with the comment history and the diffs describing the fix for the vulnerability to determine a classification. This resulted in 111 files involved with vulnerabilities. The inter-rater reliability between the two raters when classifying vulnerable files was $\kappa = 0.6$. These 111 files were then added to the 247 files found on the Mozilla Security Advisories blog, resulting in a final list of 358 source code files that were involved with vulnerabilities. Fourteen source code files were discarded and 8 source code files were changed based on the movement of code to another source code file or the removal of code from the system. Finally, a list of 271 unique files remained with at least one vulnerability.

After identifying the vulnerabilities for inclusion in the dataset, two humans then classify each vulnerability using an existing vulnerability classification scheme---the Common Weakness Enumeration (CWE) set of most commonly seen weaknesses in software\footnote{http://cwe.mitre.org/data/definitions/1003.html}. After classifying each vulnerability, the humans then convene and resolve any differences that have occurred between the two of them. If they could not come to a consensus, a third party arbitrator was used to resolve the conflict.

After this classification, the Mozilla Firefox dataset contains 14 different types of vulnerabilities and some files are associated with multiple types of vulnerabilities. To simulate vulnerability inspections targeting specific types of vulnerabilities, we grouped the vulnerabilities into five categories to ensure that each category contains enough data (\#Vulnerable Files) for a simulation, as shown in Table~\ref{tab:data}. In the following sections \tion{experiments}, we show results on simulations targeting each vulnerability category with file-level granularity. For example, when targeting ``Protection Mechanism Failure'', only the 119 files associated with this category are considered vulnerable.

\subsection{Independent Variables}
\label{sec:independent}

Here we briefly describe the three types of features extracted from the Mozilla Firefox vulnerability dataset and used in our simulations.

\subsubsection{Software Metrics}
\label{sec:metrics}

SciTools' Understand\footnote{http://www.scitools.com} was used to measure the metrics from the Mozilla Firefox source code files. The list of the software metrics provided by the dataset is shown below:
\begin{itemize}
    \item \textit{CountClassBase} -  Number of subclasses in a file.
    \item \textit{CountClassCoupled} - Coupling of the classes in a file.
    \item \textit{CountClassDerived} -   Number of subclasses derived from classes originating in a file.
    \item \textit{CountDeclInstanceVariablePrivate} -  Number of private variables declared in a     file.
    \item \textit{CountDeclMethod} - Number  of methods declared in a  file.
    \item \textit{CountInput} - Number of incoming calls to the source code file.
    \item \textit{CountOutput} -  Number of outgoing calls from the source code file.
    \item \textit{Cyclomatic} -   Cyclomatic Complexity of a file.
    \item \textit{CountLine} - Number of lines of code (excluding comments and whitespace) in a file.
    \item \textit{MaxInheritanceTree} - The size of the maximum leaf in the inheritance tree leading from a file.
\end{itemize}
Some metrics in Shin {\it et al.}~\cite{shin2011evaluating} are not covered due to the following reasons: (1) metrics that no public tool (to our knowledge) provides an equivalent to, e.g. incoming closure and outgoing closure; (2) metrics that do not apply to the Mozilla Firefox project, e.g. organization intersection factor; (3) metrics that are identical to some already covered ones in Mozilla Firefox project, e.g. edit frequency.

\subsubsection{Text Mining Features}
\label{sec:text}

As described in \tion{feature_extraction}, text mining features were extracted by tokenizing the source code files, selecting top 4000 tokens based on tf-idf score, and then applying L2 normalization on files.

\begin{table*}[!tbh]
\caption{Experiments design}
\label{tab:experiments}
\begin{center}
\setlength\tabcolsep{3.5pt}
\begin{tabular}{l|cccc}
& Target research questions  &  \makecell{Human error rate $E_R$ in \ref{step:3}} & \makecell{Use of \ref{step:5} and \ref{step:6}?} & \makecell{Recall estimation $|R_E|$ in \ref{step:7}}
\\\hline
\textbf{Target 1 efficiency} & {\bf RQ1} & $E_R=0$ & no & $|R_E| = |R|$\\
\textbf{Target 2 stopping rule} & {\bf RQ2} & $E_R=0$ & no & \makecell{$|R_E|$ is estimated by SEMI \\as described in \tion{recall_estimation}}\\
\makecell{\bf Target 3 human error\\\bf correction} & {\bf RQ3, RQ4} &  $E_R\in \{0, 0.1, 0.2, 0.3, 0.4, 0.5\}$ & yes & \makecell{$|R_E|$ is estimated by SEMI \\as described in \tion{recall_estimation}}\\
\hline
\end{tabular}
\end{center}
\end{table*}

\subsubsection{Crash Features}
\label{sec:crash}

Crash dump stack trace data was collected from Mozilla Crash Reports\footnote{https://crash-stats.mozilla.com/home/product/Firefox}. We collected crashes from January 2017 to November 2017, with a total of 1,141,519 crashes collected. For each crash, the field marked ``crashing thread'' is observed and each file that appeared in the thread is added to the dataset. We also kept track of how many times a file was observed in different crashes since files that crash more often have a higher chance to contain vulnerabilities.

\subsection{Simulation on Mozilla Firefox Dataset}

Code inspection with {\IT}, as described in \tion{methodology}, is simulated on the Mozilla Firefox dataset with the following settings:

\begin{enumerate}[start=1,label={}]
\item
\textbf{Step 1 Feature Extraction:} $E$ is the set of all 28,750 source code files, $R$ is the set of source code files containing the target type of vulnerabilities shown in Table~\ref{tab:data}, $L \leftarrow \emptyset$ is the set of labeled files, and $L_R \leftarrow \emptyset$ is the set of labeled target vulnerable files. Extract \textbf{features} from each source code file in $E$.
\item
\textbf{Step 2 Initial Sampling: }The batch size is chosen as $N_1=100$. We use $N_1=100$ during our simulations to speed up the simulation process. However, it is suggested to use a smaller batch size, e.g. $N_1=10$, in practice since the cost of training a model more often is usually much less than the cost of inspecting more code.
\item
\textbf{Step 3 Human Oracle: } Rather than asking human inspectors to inspect source code files, as specified in \tion{human_oracle}, the ground truth labels from Mozilla Firefox dataset are applied to simulate the inspection. Each vulnerable file $x\in Q\cap R$ has $E_R$ chance of being wrongly labeled as ``non-vulnerable'' and $1-E_R$ chance of being correctly labeled as ``vulnerable''. Every non-vulnerable file $x\in Q\setminus R$ will be labeled as ``non-vulnerable''. Here $E_R=0,10,20,30,40,50\%$ simulates the probability of a human expert failing to detect the vulnerabilities (false negative rate).
\item
\textbf{Step 5 Error Prediction: } Half of the inspected files are selected for double checks ($N_2=0.5 N_1$) based on the DISPUTE principle. We show later in \tion{experiments} that by double-checking half of the labeled files, most (96\%) of the human false negatives can be covered.
\item
\textbf{Step 8 Query Strategy: }Uncertainty sampling if $|L_R|< N_3 = 10$; certainty sampling when $|L_R|\ge N_3 = 10$. 
\end{enumerate}
Here, $N_1, N_2, N_3$ are selected based on our experience in solving other total recall problems~\cite{Yu2018,Yu2019}.

\section{Experiments and Results}\label{sec:experiments}

This section describes experiments assessing {\IT}'s performance. All the following experiments are simulated on the Mozilla Firefox case study, as described in \tion{case-study} and are used to answer the research questions listed in \tion{introduction}.

\subsection{Performance Metrics}\label{sec:performance Metrics}

{\IT}'s task is to optimize the inspection effort for achieving very high recall. Therefore {\IT} focuses on minimal cost, maximal recall, defined as follows:
\begin{equation}\label{eq:recall}
\mathit{recall}  = \frac{\#\,\mathit{of\,vulnerable\,files\,found}}{\#\,\mathit{of\,vulnerable\,files\,exist}} = \frac{|L_R|}{|R|}.
\end{equation}
\begin{equation}\label{eq:cost}
\mathit{cost} = \frac{\#\,\mathit{of\,source\,code\,files\,reviewed}}{\#\,\mathit{of\,source\,code\,files\, exist}} = \frac{|L|}{|E|}.
\end{equation}
Since our simulations are in file level granularity, the numerator of cost in \eqref{eq:cost} counts the number of times the source code files being reviewed, i.e. it still increases when the same file is reviewed for a second time by a different reviewer. Using these two metrics, one treatment is considered better than another if it reaches the same target recall with a lower cost.

Beside recall and cost, we use the Estimation vs Cost curve to assess the accuracy for estimating the total number of vulnerable files in \ref{step:7}, \fig{{\IT}}:
\begin{equation}\label{eq:est}
\mathit{estimation}  = \frac{\#\,\mathit{of\,vulnerable\,files\,estimated}}{ \#\,\mathit{of\,vulnerable\,files\,exist}} = \frac{|R_E|}{|R|}.
\end{equation}
The sooner this estimation converges to 1.0, the better the estimator is. As described in \ref{step:7}, the inspection process will stop when $$|L_R| \ge  T_{rec}|R_E|,$$ where $T_{rec}$ represents the target recall. The closer the inspection stops to the target recall, the better the stopping rule.

As shown in Table~\ref{tab:experiments}, three experiments are designed to answer the research questions listed in \tion{introduction} and evaluate {\IT} based on the three targets described in \tion{total_recall_solution}. More details on each experiment's design and its corresponding result are presented in the rest of this section.

\subsection{Target 1 Efficiency}
\label{sec:exp_efficiency}

\subsubsection{Simulation Design}

Cost for reaching different levels of recall is used to assess the vulnerability inspection efficiency of {\IT}. As a result, stopping rule and human errors are not considered, i.e. human error rate $E_R$ in \ref{step:3} is set to be $0$, \ref{step:5} and \ref{step:6} are disabled, and in \ref{step:7}, real recall is used for the stopping rule $|R_E| = |R|$, as shown in Table~\ref{tab:experiments}.

We first need to decide (1) what feature in \ref{step:1} serves best for predicting vulnerabilities in the active learning-based framework, and (2) what type of learner (supervised, unsupervised, semi-supervised) in \ref{step:4} performs best inside the active learning-based framework. As a result, we first test the following feature types with supervised learner (linear SVM as described in \tion{feature_extraction}) to find the best feature set:
\begin{itemize}
\item
\textbf{Metrics}: a linear SVM trained on file level software metrics features (described in \tion{metrics}) which quantifies different types of software complexity and are collected from the source code.
\item
\textbf{Text}:  a linear SVM trained on file level text mining features (described in \tion{text}) which treats the source code as raw text and performs standard text mining feature extraction (term frequency with L2 normalization~\cite{Yu2018}) on it.
\item
\textbf{Hybrid}: a linear SVM trained on the combination of text mining features and crash features (described in \tion{crash}). This feature set is built  by adding one column (number of times the source code file has crashed) to the term frequency matrix of the text mining features set before normalization. In \ref{step:2}, crash counts are applied as domain knowledge to first sample files which have crashed most frequently.
\end{itemize}
Then we test the following semi-supervised and unsupervised learners within the active learning-based framework:
\begin{itemize}
\item
\textbf{S3VM}: a linear kernel QN S3VM trained on file level text mining features, similar to \textbf{Text} except that it utilizes unlabeled data in training to build its model.
\item
\textbf{Crash}: an unsupervised approach where no model is trained and source code files are selected in descending order of the number of times they crashed~\cite{theisen2015strengthening} (described in \tion{crash}). Note that this unsupervised approach does not require the active learning-based framework of {\IT}.
\end{itemize}
As a baseline, the simulation also includes the following treatment:
\begin{itemize}
\item
\textbf{Random}: a baseline approach where source code files are inspected in random order. This treatment serves as a baseline where code inspection is performed without any help from {\IT}.
\end{itemize}
Summaries of the treatments are provided in Table~\ref{tab:treatments}. {\IT} framework includes all the treatments learned from human oracles.

\begin{table}[!tb]
\caption{Treatments for \textbf{Target 1 Efficiency}}
\label{tab:treatments}
\begin{tabular}{l|c|cccc|}
\multicolumn{2}{c|}{} & \multicolumn{4}{c|}{\textbf{Learn From}} \\\cline{3-6}
\textbf{Treatment} & \textbf{Learner} &  \makecell{software \\metrics} & \makecell{text mining \\features} & \makecell{crash \\features} & \makecell{human \\oracle} \\ \hline
Metrics & SVM & \checkmark & & & \checkmark \\
Text & SVM & & \checkmark  & & \checkmark \\
Hybrid & SVM & & \checkmark & \checkmark & \checkmark \\\hline
S3VM & S3VM & & \checkmark  & & \checkmark \\\hline
Crash & n/a & & & \checkmark & \\\hline
Random & n/a & & & & \\\hline
\end{tabular}
\end{table}

\begin{table*}[!tb]
\caption{Experimental Results for Target 1 Efficiency}
\label{tab:recall}
\begin{center}
\begin{threeparttable}
\scriptsize
\setlength\tabcolsep{16pt}
\begin{tabular}{l|l|r@{ (}c@{) }|r@{ (}c@{) }|r@{ (}c@{) }|r@{ (}c@{) }|r@{ (}c@{) }|r@{ (}c@{) }|r@{ (}c@{) }|r@{ (}c@{) }| }
&        & \multicolumn{16}{c|}{Target recall}\\\cline{3-18}
&        & \multicolumn{2}{c|}{60} & \multicolumn{2}{c|}{70} & \multicolumn{2}{c|}{80} & \multicolumn{2}{c|}{85} & \multicolumn{2}{c|}{90} & \multicolumn{2}{c|}{95} & \multicolumn{2}{c|}{99} & \multicolumn{2}{c|}{100}\\\cline{3-18}
&Vulnerability Type& \multicolumn{16}{c|}{Cost to reach target recall}\\\hline
\multirow{7}{*}{Metrics}  & Protection Mechanism Failure                                      & 50 &7  & 56 &6  & 60 &8  & 63 &8  & 67 &8  & 79 &7 & 94 &3  & 99 &2  \\
 & Resource Management Errors & 49 &25 & 57 &15 & 63 &13 & 67 &12 & 71 &10 & 75 &9 & 99 &0  & 99 &0  \\
 & Data Processing Errors                                         & 35 &4  & 41 &8  & 45 &9  & 61 &23 & 70 &24 & 78 &9 & 82 &5  & 82 &5  \\
 & Code Quality                                        & 38 &6  & 42 &20 & 61 &19 & 61 &18 & 64 &12 & 66 &9 & 72 &3  & 72 &3  \\
  & Other &	38&19&	45&17&	58&7&	60&6&	64&8&	70&10&	92&12&	92&12  \\
 & All                                                 & 49 &17 & 53 &7  & 60 &4  & 63 &4  & 66 &3  & 72 &3 & 94 &6  & 99 &0  \\
 \cellcolor{white}  & \cellcolor{gray!25} Median &  42 & 18 &	50 & 17	& 59 & 12 &	62 & 10	& 67 & 8	& 74 & 10	& 91 & 15	& 94 & 19 \\
\hline
\multirow{7}{*}{Text}   & Protection Mechanism Failure                                      & 6 &1  & 9 &2  & 12 &1 & 16 &1 & 20 &3 & 28 &3 & 37 &9  & 43 &7  \\
 & Resource Management Errors & 6 &1  & 8 &2  & 9 &1  & 9 &2  & 11 &3 & 13 &2 & 33 &4  & 33 &4  \\
 & Data Processing Errors                                         & 12 &4 & 13 &2 & 14 &4 & 15 &4 & 16 &4 & 20 &4 & 42 &15 & 42 &15 \\
 & Code Quality                                        & 4 &4  & 5 &4  & 8 &3  & 10 &4 & 19 &5 & 21 &5 & 30 &7  & 30 &7  \\
   & Other&	5&3&	6&4&	11&5	&12&5	&14&4&	18&3&	33&2&	83&2  \\
 & All                                                 & 5 &0  & 7 &0  & 9 &1  & 12 &1 & 14 &1 & 20 &2 & 34 &2  & 85 &0  \\
 \cellcolor{white}  & \cellcolor{gray!25} Median   & 6&2	&8&3&	10&4&	13&4&	16&5&	20&7&	34&7	&43&49 \\
\hline
\multirow{7}{*}{Hybrid}  & Protection Mechanism Failure                                      & 7 &0 & 9 &0  & 12 &0 & 14 &0 & 16 &1 & 21 &3  & 58 &1 & 59 &1  \\
 & Resource Management Errors & 5 &0 & 7 &0  & 8 &0  & 9 &0  & 11 &0 & 13 &2  & 39 &0 & 39 &0  \\
 & Data Processing Errors                                         & 7 &0 & 10 &0 & 15 &0 & 15 &0 & 16 &0 & 29 &2  & 37 &3 & 37 &3  \\
 & Code Quality                                        & 3 &1 & 4 &1  & 4 &2  & 5 &1  & 10 &0 & 12 &0  & 16 &1 & 16 &1  \\
 & Other &	4&0&	9&0&	11&0&	12&0&	14&0&	14&0&	37&9&	61&14\\
 & All                                                 & 6 &0 & 8 &0  & 10 &0 & 11 &0 & 14 &0 & 18 &2  & 36 &0 & 59 &0  \\
 \cellcolor{white} & \cellcolor{gray!25} Median   & 6&2&	8&1&	10&4&	12&5&	14&4&	17&7&	37&5&	41&22 \\
\hline
\multirow{7}{*}{S3VM}  & Protection Mechanism Failure                                        & 20&44 & 22&43 & 25&45 & 28&41 & 33&37 & 52&23 & 62&16 & 70&15 \\
& Resource Management Errors & 21&22 & 23&21 & 24&22 & 25&21 & 30&20 & 32&18 & 51&8  & 51&8  \\
& Data Processing Error                                    & 25&37 & 26&38 & 28&40 & 29&39 & 32&36 & 37&29 & 43&28 & 43&28 \\
&Code Quality                                        & 29&20 & 29&19 & 30&19 & 30&19 & 35&26 & 41&23 & 53&21 & 53&21 \\
&Other                                               & 28&20 & 29&21 & 30&19 & 32&17 & 33&17 & 36&17 & 53&40 & 53&40    \\
&All                                                 & 20&8  & 21&8  & 25&7  & 27&7  & 31&5  & 41&5  & 64&10 & 80&7  \\
 \cellcolor{white}  & \cellcolor{gray!25} Median   & 23&24 & 25&23 & 27&22 & 29&20 & 32&22 & 40&23 & 56&27 & 60&35 \\ \hline
\multirow{7}{*}{Crash} & Protection Mechanism Failure                                      & 5 &0 & 8 &0  & 13 &0 & \multicolumn{2}{c|}{n/a}  & \multicolumn{2}{c|}{n/a}  & \multicolumn{2}{c|}{n/a}  & \multicolumn{2}{c|}{n/a}  & \multicolumn{2}{c|}{n/a}  \\
 & Resource Management Errors & 4 &0 & 5 &0  & 8 &0  & 10 &0 & 11 &0 & \multicolumn{2}{c|}{n/a}  & \multicolumn{2}{c|}{n/a}  & \multicolumn{2}{c|}{n/a}  \\
 & Data Processing Errors                                         & 8 &0 & 11 &0 & \multicolumn{2}{c|}{n/a}  & \multicolumn{2}{c|}{n/a}  & \multicolumn{2}{c|}{n/a}  & \multicolumn{2}{c|}{n/a}  & \multicolumn{2}{c|}{n/a}  & \multicolumn{2}{c|}{n/a}  \\
 & Code Quality                                        & 1 &0 & 2 &0  & 4 &0  & 4 &0  & 10 &0 & 11 &0 & 12 &0 & 12 &0 \\
 & Other&	6&0&	9&0&	12&0&	\multicolumn{2}{c|}{n/a}	&\multicolumn{2}{c|}{n/a}	&\multicolumn{2}{c|}{n/a}	&\multicolumn{2}{c|}{n/a}&	\multicolumn{2}{c|}{n/a}  \\
 & All                                                 & 5 &0 & 8 &0  & 11 &0 & 14 &0 & \multicolumn{2}{c|}{n/a}  & \multicolumn{2}{c|}{n/a}  & \multicolumn{2}{c|}{n/a}  & \multicolumn{2}{c|}{n/a}  \\
 \cellcolor{white}  & \cellcolor{gray!25} Median    & 5&1	&8&3	&11&4  & \multicolumn{2}{c|}{n/a} & \multicolumn{2}{c|}{n/a} & \multicolumn{2}{c|}{n/a} & \multicolumn{2}{c|}{n/a} & \multicolumn{2}{c|}{n/a}\\
\hline
\multirow{7}{*}{Random}  & Protection Mechanism Failure                                      & 59 &7  & 70 &4  & 80 &4  & 85 &3 & 90 &2  & 95 &2 & 98 &1 & 99 &0 \\
 & Resource Management Errors & 59 &4  & 69 &5  & 77 &5  & 84 &4 & 88 &3  & 94 &2 & 99 &1 & 99 &1 \\
 & Data Processing Errors                                         & 57 &12 & 70 &11 & 78 &10 & 84 &9 & 86 &10 & 94 &5 & 97 &3 & 97 &3 \\
 & Code Quality                                        & 58 &8  & 70 &10 & 77 &11 & 82 &8 & 90 &9  & 94 &6 & 98 &2 & 98 &2 \\
  & Other&	59&5&	67&5&	78&6&	83&5&	88&6&	93&5	&97&2&	97&2\\
 & All                                                 & 59  &3  & 69 &3  & 79 &3  & 84 &2 & 89 &2  & 94 &1 & 98 &0 & 99 &0 \\
 \cellcolor{white}  & \cellcolor{gray!25} Median           & 59&6	&69&6&	79&5&	83&5&	89&4&	94&3&	98&1&	98&2 \\
\hline
\end{tabular}
\begin{tablenotes}\small
This table shows the cost (i.e. percent code reviewed) required to reach different levels of recall. Medians and IQRs (75th-25th percentile) are shown for 30 repeated simulations (IQR results are shown in brackets).
{\bf Crash} has many empty cells since it provides no information on vulnerabilities located in the files that have never crashed.
In this table, {\em lower} median values are {\em better} so all the {\bf Random} results are worse than anything else (as might be expected). Of the remaining results, {\bf Text} and {\bf Hybrid} perform
better than {\bf Metrics}. Important note: this table does not consider how to stop at the target recall, it only shows the cost when first reaching that recall. For experiments considering stopping rules, see Table~\ref{tab:est}.
\end{tablenotes}
\end{threeparttable}
\end{center}
\end{table*}

\subsubsection{Experimental Result}

\begin{figure*}[!t]

\begin{center}   \includegraphics[width=.9\linewidth]{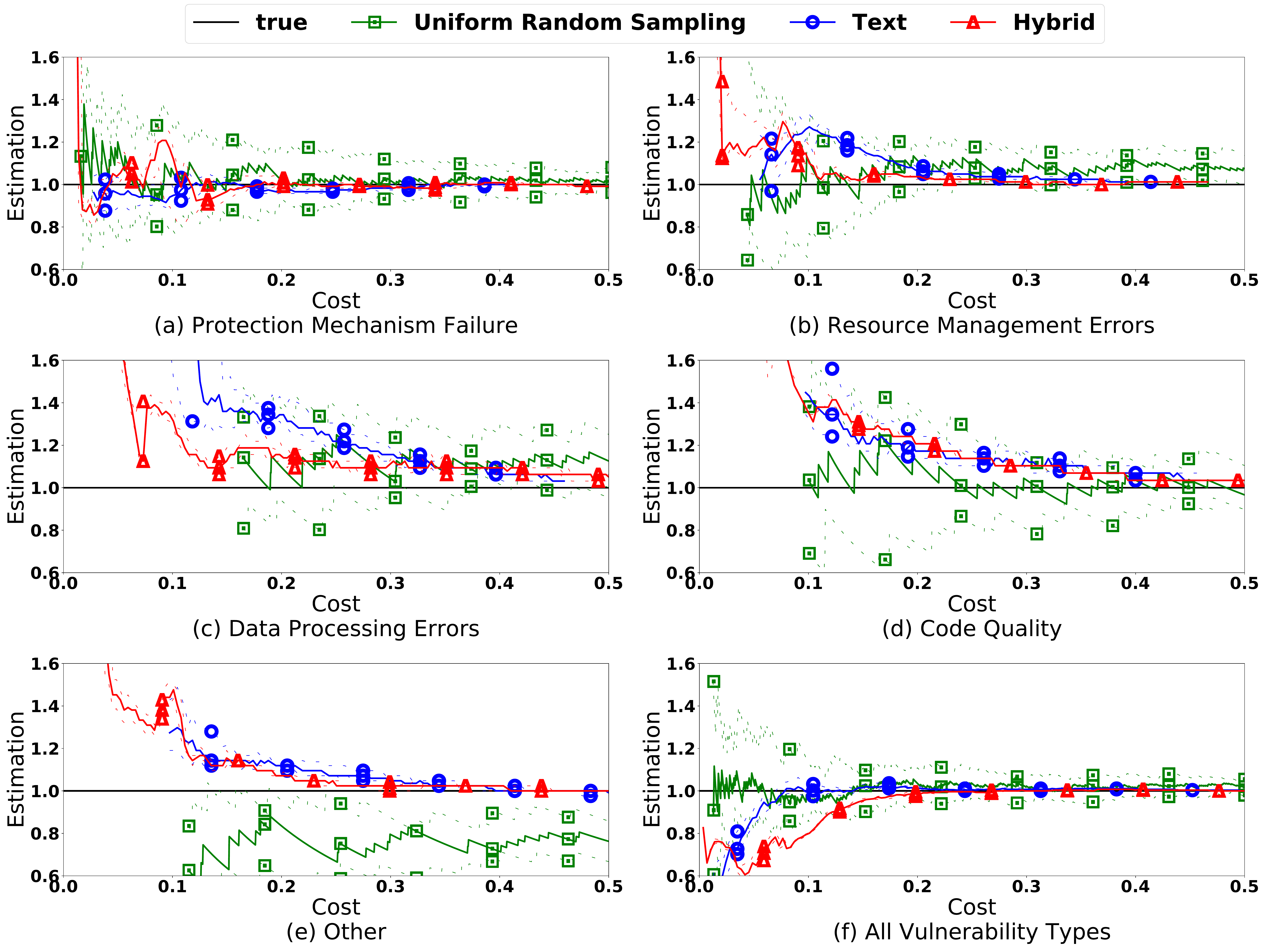}
\end{center}
\caption{Estimation vs Cost curves.
Estimations (calculated as Equation~\eqref{eq:est})  are considered accurate and useful if they  converge to estimation=1.0 early (when cost is small).
{\bf Solid lines} represent the median estimations from 30 simulations while {\bf dashed lines}
show the  75th to 25th percentile range.
In this figure, estimation=1.0 is denoted as the ``true'' line.  When an estimation converges to this line, it means that the estimated number of vulnerabilities are equal to the true number. Also, when  cost reaches 1.0, all source code files have been reviewed. For example, the estimations from \textbf{Text} and \textbf{Hybrid} on the entire project converge to 1.0 when cost is about 0.25, thus providing accurate estimation on whether the target recall has reached for cost $\ge$ 0.25. }
\label{fig:est}
   
\end{figure*}

{\bf RQ1: Can human inspection effort be saved by applying {\IT} to find a certain percentage of vulnerabilities?} Table~\ref{tab:recall} shows the cost to reach different levels of recall. One method is considered better than another if it costs less to reach the same recall. Based on these results, we make the following observations:
\begin{itemize}
\item
\textbf{Metrics}, \textbf{Text}, \textbf{Hybrid}, and \textbf{S3VM} all perform better than \textbf{Random}, indicating the effectiveness of the proposed active learning framework in {\IT}.
\item
\textbf{Crash} performs well in the early stages (when recall$<$80\%) but is unable to achieve certain target recall values (see the numerous ``n/a'' entries of Table~\ref{tab:recall}). For vulnerabilities of type ``Code Quality'', \textbf{Crash} performs better than any other approach. Overall, this unsupervised approach has limitations compared to active learning-based approaches.
\item
Targeting different types of vulnerabilities does not affect much of the performance of {\IT} with active learning.
\end{itemize}
Therefore active learning-based {\IT} approach is recommended to reliably achieve high recall with low cost. Among the active learning-based approaches:
\bi
\item
Software metrics features (\textbf{Metrics}) perform the worst---they always cost more to reach the same recall. Therefore we do not recommend using static software metrics as features to predict vulnerabilities in {\IT}.
\item
When crash features are unavailable, e.g. before a software's first release, \textbf{Text} performs the best.
\item
When crash features are available, \textbf{Hybrid} utilizes both \textbf{Text} and \textbf{Crash} and performs slightly better than \textbf{Text} in terms of median performances and greatly reduces the variance.
\item
A semi-supervised learner with text mining features (\textbf{S3VM}) performs worse than its supervised counterpart (\textbf{Text}), especially in the early stages when training data is few. Therefore semi-supervised learners are not recommended inside the active learning-based framework.
\item
These results also show that the human effort required is still high even with the help of {\IT}, e.g. to identify 271 $\times$ 0.95 = 257 vulnerable files, 28,750 $\times$ 0.2 = 5,750 files need to be inspected by humans. While this means 28,750 - 5,750 = 23,000 fewer files to be inspected compared to inspecting all the files, it is still a tedious and frustrating task (one vulnerable file may be found within every 22 files inspected).
\ei
Summing up this work on {\bf RQ1}, we say:

\begin{RQ}{RQ1: Can human inspection effort be saved by applying {\IT} to find a certain percentage of vulnerabilities?}
Yes. Simulated on Mozilla Firefox dataset, applying {\IT} with linear SVM trained on text mining features, 60, 70, 80, 90, 95, 99\% of the target vulnerabilities can be found by inspecting only 6, 8, 10, 16, 20, 34\% of the source code files, respectively. If available, crash features can further boost the inspection efficiency when applied to guide the inspection order in the early stage. 
\end{RQ}

\subsection{Target 2 Stopping Rule}
\label{sec:stopping_rule}

\subsubsection{Simulation Design}

In this experiment,
the SEMI estimator
(described in \tion{recall_estimation}), is applied to estimate the number of vulnerabilities when running {\IT} with the best feature set picked from \textbf{RQ1}. The estimation from SEMI is then used as an indicator of whether the target recall has been reached and thus the inspection can be safely stopped. This experiment is designed to test the accuracy of the SEMI estimations and whether the inspections can stop close to the target recall. \textbf{Uniform random sampling} is applied as a baseline estimation. \textbf{Uniform random sampling} samples from unlabeled data randomly and queries the labels, then estimates $|R|$ as $|R_E|=|E|\times|L_R|/|L|$. It was believed to be the most accurate estimation method before the introduction of SEMI~\cite{wallace2013active}. 

Human errors are considered in the next experiment so, for this experiment, the  human error rate $E_R$ in \ref{step:3} is set to be $0$, and \ref{step:5} and \ref{step:6} are disabled, as shown in Table~\ref{tab:experiments}.  

\subsubsection{Experimental Result}

{\bf RQ2: Can {\IT} correctly stop the vulnerability inspection when a predetermined percentage of vulnerabilities has been found?} To answer this research question, we first show the accuracy of the estimations from SEMI by presenting the Estimation vs Cost curve in \fig{est}.
In these plots, estimations, calculated as Equation~\eqref{eq:est}, are considered accurate and useful if they  converge to estimation=1.0 (denoted as the ``true'' line) early (when cost is small).
According to \fig{est}, the estimations from SEMI using either \textbf{Hybrid} or \textbf{Text} feature converge to 1.0 earlier than \textbf{Uniform random sampling} on every target vulnerability types. Usually the estimation error of SEMI becomes $\le$ 5\% after cost $\ge$ 0.3.

Table~\ref{tab:est} shows results  using 
the SEMI stopping rule.
In most cases, the SEMI estimator slightly over-estimates ($|R_E|>|R|$) to ensure the target recall is reached.
Also, the higher the target recall, the more effective  the stopping rule (i.e. estimation becomes     more accurate when cost increases).
Further, the stopping rule works better with \textbf{Text}, where the inspection usually stops with less than 6\% error from target recall. Finally, \textbf{Hybrid} over-estimates ($|R_E|>|R|$) too much on ``Code Quality'' and under-estimates ($|R_E|<|R|$) on ``All''. Overall, when using SEMI, \textbf{Text} is better.

\begin{table}
\caption{Experiment Results for Target 2 Stopping Rule}
\label{tab:est}
\begin{center}
\begin{threeparttable}
\scriptsize
\setlength\tabcolsep{3.5pt}
\begin{tabular}{l|l|r@{ (}c@{) }|r@{ (}c@{) }|r@{ (}c@{) }|r@{ (}c@{) }|r@{ (}c@{) }|r@{ (}c@{) }|}
                         \multicolumn{2}{c|}{~}                & \multicolumn{12}{c|}{target recall}\\\cline{3-14}
                         \multicolumn{2}{c|}{~}       & \multicolumn{4}{c|}{90} & \multicolumn{4}{c|}{95} &
                         \multicolumn{4}{c|}{99} \\ \cline{3-14}
                          & \makecell{Vulnerability Type}                  & \multicolumn{2}{c|}{Recall}     & \multicolumn{2}{c|}{Cost}       & \multicolumn{2}{c|}{Recall}     & \multicolumn{2}{c|}{Cost}       & \multicolumn{2}{c|}{Recall}     & \multicolumn{2}{c|}{Cost}       \\ \hline
\parbox[t]{3mm}{\multirow{7}{*}{\rotatebox[origin=c]{90}{Text}}}    & \makecell[l]{Protection Mechanism\\ Failure}                                      & 87&2 & 18&2  & 91&2  & 22&3  & 98&3  & 33&5  \\
 & \makecell[l]{Resource Management\\ Errors} & 96&0 & 19&2  & 98&1  & 26&3  & 100&0 & 46&3  \\
 & Data Processing Errors                                         & 96&2 & 36&5  & 100&3 & 46&5  & 100&0 & 56&6  \\
 & Code Quality                                        & 96&3 & 30&7  & 100&0 & 42&6  & 100&0 & 54&7  \\
 & Other&	97&0&	25&6&	97&0&	32&6	&97&0	&50&5 \\
 & All                                                 & 90&2 & 15&2  & 95&1  & 21&1  & 99&0  & 43&0  \\
 \cellcolor{white}  & \cellcolor{gray!25} Median                                              & 96&6	&23&12	&97&4	&30&17	&99&2	&47&10\\ \hline
 \parbox[t]{3mm}{\multirow{7}{*}{\rotatebox[origin=c]{90}{Hybrid}}}   & \makecell[l]{Protection Mechanism\\ Failure}                                      & 84&19 & 14&7  & 94&2  & 20&6  & 98&1  & 44&12 \\
 & \makecell[l]{Resource Management\\ Errors} & 93&1  & 12&0  & 98&1  & 17&1  & 100&0 & 48&0  \\
 & Data Processing Errors                                         & 96&3  & 31&3  & 100&0 & 45&2  & 100&0 & 58&2  \\
 & Code Quality                                        & 100&0 & 27&0  & 100&0 & 38&0  & 100&0 & 58&1  \\
 & Other&	95&0&	20&0	&95&0	&28&1&	97&0&	52&3\\
 & All                                                 & 67&2  & 8&0   & 70&2  & 8&0   & 99&0  & 46&0  \\
\cellcolor{white}  & \cellcolor{gray!25} Median                                              & 93&11	&20&14&	96&5	&27&21	&100&1	&50&10  \\ \hline
\end{tabular}
\begin{tablenotes}\small
This table shows the actual achieved recalls and cost when using the SEMI estimator to stop at target recall.
Same as in Table~\ref{tab:recall}, numbers in brackets denote IQR values (75th-25th percentile ranges) and the other numbers are median values across 30 repeated simulations.
\end{tablenotes}
\end{threeparttable}
\end{center}
\end{table}

\begin{RQ}{RQ2: Can {\IT} correctly stop the vulnerability inspection when a predetermined percentage of vulnerabilities has been found?}
Yes. Through accurate estimation of the number of vulnerabilities, {\IT} can stop the inspection process close to the target recall. We recommend using text mining features and setting the target recall as high as 95\% or 99\% to guarantee the detection of most vulnerabilities.
\end{RQ}

The above results also suggest that it is very important to further improve the estimation since a tiny reduction in its error can lead to the saving of a large amount of effort. For example, when targeting 95\% recall, the inspection stopped at 97\% recall with 30\% cost, which means 97\% of the vulnerable files can be identified by inspecting 8,625 source code files. Compared to the result of RQ1, about 3,000 more files need to be inspected due to the small estimation error (2\%) of {\IT}. 

\subsection{Target 3 Human Error Correction}
\label{sec:errors}

\subsubsection{Simulation Design}
Hatton~\cite{hatton2008testing}'s report suggested that each  ``vulnerable'' file has in average 47\% chance of being mislabeled as ``non-vulnerable'' while ``non-vulnerable'' files can never be mislabeled as ``vulnerable''. Following this report, we simulate how vulnerability inspection would be affected by randomly injecting human errors:
\bi
\item
No false positives: ``non-vulnerable'' files can never be mislabeled as ``vulnerable''.
\item
False negative rate $E_R$: each  ``vulnerable'' file has the same chance of being mislabeled as ``non-vulnerable''.
\ei
Our simulations explore five ways to handle
 human errors while injecting
 false negatives at the rate of  $E_R = 0,10,20,30,40,50\%$):
\bi
\item
\textbf{None}: No error correction. This will be our baseline result.
\item
\textbf{Two-persons~\cite{hatton2008testing}} check each label. In \ref{step:3}, if one file is labeled as ``non-vulnerable'' and has been reviewed by only one reviewer, it goes back into the queue to wait for a different reviewer to double-check it.
\item
\textbf{Cormack'17~\cite{Cormack2017Navigating}}: This is an advanced error correction method for citation screening. We reproduce half of Cormack'17 to correct false negatives only and incorporate it into the distributed framework:
In \ref{step:8}, Cormack'17 utilizes a different stopping rule. This method detects the inflection point\footnote{The inflection (or ``elbow'') point 
has the longest distance to the  line between the 
recall starting point and endpoint~\cite{Cormack2016Engineering}.} $i$ of the current recall curve, and compare the slopes before and after $i$. If $slope_{<i}/slope_{>i}$ is greater than a specific threshold $\rho = 6$, the review should be stopped. For details about this stopping rule, please refer to Cormack and Grossman~\cite{Cormack2016Engineering}. After the stopping rule is satisfied in \ref{step:8}, send all files reviewed before the inflection point $i$ and labeled as ``non-vulnerable'' to the queue. The security review and test stops after all the files in the queue have been reviewed by a different reviewer.
\item
\textbf{DISPUTE}: As described in \tion{error_prediction}, every iteration, $M=0.5N$ of the labeled ``non-vulnerable'' files are selected based on how much the active learner disagrees with their current labels. It then pushes the selected files back into the queue and asks a different human expert for double-checking. In simulations, these double checks have the same false negative rate $E_R$.
\item
\textbf{DISPUTE(3)}: Similar to DISPUTE but selected files are inspected by \emph{two}  humans (so ``vulnerable'' files are less likely to be missed again albeit doubling the double-checking cost).
\ei

\subsubsection{Experimental Result}

\begin{table*}
\caption{Experimental Results for Target 3 Human Error Correction}
{\scriptsize
\begin{center}
\label{tab:error}
{
\setlength\tabcolsep{4pt}
\begin{threeparttable}
\begin{tabular}{c|l| r@{ (}c@{) } r@{ (}c@{) } r@{ (}c@{) } r@{ (}c@{) } r@{ (}c@{) }| r@{ (}c@{) } r@{ (}c@{) } r@{ (}c@{) } r@{ (}c@{) } r@{ (}c@{) }}
Error &  & \multicolumn{10}{c|}{Relative Recall = }& \multicolumn{10}{c}{Relative Cost = }\\
  Rate & Vulnerability Type & \multicolumn{10}{c|}{Observed Recall / Baseline Recall (from Table~\ref{tab:est})}& \multicolumn{10}{c}{Observed Cost / Baseline Cost  (from Table~\ref{tab:est})}\\\cline{3-22}
($E_R$) &  & \multicolumn{2}{m{1.1cm}}{None}   & \multicolumn{2}{m{1.2cm}}{Two-person} &  \multicolumn{2}{m{1.1cm}}{Cormack'17} & \multicolumn{2}{m{1.0cm}}{DISPUTE} & \multicolumn{2}{m{1.2cm}|}{DISPUTE(3)}  & \multicolumn{2}{m{1.1cm}}{None}   & \multicolumn{2}{m{1.2cm}}{Two-person} &  \multicolumn{2}{m{1.1cm}}{Cormack'17} & \multicolumn{2}{m{1.0cm}}{DISPUTE} & \multicolumn{2}{m{1.2cm}}{DISPUTE(3)}\\\hline
\multirow{7}{*}{0\%} & Protection Mechanism Failure                                      & 100&1  & 100&1  & 91&8  & 98&2   & 98&2   & 100&11  & 200&22  & 60&22  & 128&27  & 171&36  \\
 & Resource Management Errors & 100&0  & 100&0  & 96&0  & 100&0  & 100&0  & 100&5   & 200&10  & 56&13  & 149&7   & 198&9   \\
 & Data Processing Errors                                         & 100&0  & 100&0  & 96&0  & 100&0  & 100&0  & 99&12   & 199&24  & 69&21  & 137&13  & 181&17  \\
 & Code Quality                                        & 100&0  & 100&0  & 89&6  & 100&0  & 100&0  & 100&6   & 200&12  & 43&13  & 145&8   & 192&12  \\
 & Other&	100&0 &	100&0 &	78&17 &	100&0 &	100&0 &	100&10 &	200&20 	&43&34 &	144&11 &	190&10 \\
 & All                                                 & 100&0  & 100&0  & 95&0  & 100&0  & 100&0  & 100&0   & 200&0   & 47&6   & 150&1   & 201&2   \\
\cellcolor{white}        & \cellcolor{gray!25} Median                                              & 100&0 &	100&0 &	95&7 	&100&0 	&100&0 	&100&6 &	200&13 	&52&22 	&146&13 &	192&19   \\\hline
\multirow{7}{*}{10\%}  & Protection Mechanism Failure                                      & 90&3  & 99&1   & 87&12  & 96&2   & 97&2   & 97&8    & 199&26  & 58&31  & 116&14  & 148&35  \\
 & Resource Management Errors & 91&4  & 98&1   & 95&1   & 99&1   & 99&1   & 102&11  & 199&15  & 57&16  & 147&14  & 197&19  \\
 & Data Processing Errors                                         & 90&3  & 100&3  & 90&14  & 98&3   & 100&2  & 105&11  & 201&24  & 61&30  & 140&15  & 180&28  \\
 & Code Quality                                        & 91&6  & 98&3   & 86&5   & 100&2  & 100&0  & 101&8   & 200&17  & 46&28  & 151&10  & 195&7   \\
 & Other&	92&4 &	100&0 &	92&12 &	100&1 &	100&0 &	101&8 &	200&16 	&59&44 &	145&10 &	191&8 \\
 & All                                                 & 90&2  & 99&0   & 92&1   & 98&0   & 100&0  & 99&1    & 200&2   & 49&4   & 149&2   & 200&1   \\
\cellcolor{white}  & \cellcolor{gray!25} Median                                              & 90&4 	&99&2 &	91&9 	&98&2 	&100&1 &	100&9 	&200&15 &	54&23 	&146&14 &	193&21   \\\hline
\multirow{7}{*}{20\%}  & Protection Mechanism Failure                                      & 80&1   & 95&2  & 82&9   & 96&4  & 97&3  & 103&11  & 196&26  & 63&26  & 130&23  & 155&37  \\
 & Resource Management Errors & 79&2   & 96&2  & 91&5   & 95&2  & 97&2  & 105&11  & 204&22  & 60&4   & 147&15  & 196&19  \\
 & Data Processing Errors                                         & 78&9   & 93&5  & 87&28  & 93&3  & 98&3  & 102&9   & 204&28  & 64&38  & 140&15  & 180&20  \\
 & Code Quality                                        & 86&12  & 96&3  & 79&17  & 96&6  & 98&3  & 105&5   & 210&19  & 39&15  & 154&17  & 197&5   \\
 & Other&	82&7 &	97&4 &	86&20 &	96&6 &	100&0 &	98&11 &	202&26 	&49&41 &	147&14 &	191&13 \\
 & All                                                 & 79&2   & 96&1  & 87&2   & 95&1  & 99&0  & 97&2    & 199&3   & 49&7   & 148&4   & 200&3   \\
\cellcolor{white}  & \cellcolor{gray!25} Median                                              & 80&6 &	96&3 	&86&11 &	95&3 &	98&3 &	101&11 &	201&22 &	52&28 &	145&15 &	194&19  \\\hline
\multirow{7}{*}{30\%}   & Protection Mechanism Failure                                      & 68&2  & 91&4   & 65&17  & 88&3  & 93&3  & 101&4   & 201&26  & 40&43  & 116&18  & 160&38  \\
 & Resource Management Errors & 74&8  & 90&4   & 84&5   & 91&5  & 96&2  & 104&9   & 199&14  & 69&20  & 147&14  & 209&22  \\
 & Data Processing Errors                                         & 75&6  & 90&11  & 84&7   & 90&3  & 95&6  & 105&9   & 204&21  & 59&21  & 148&12  & 179&22  \\
 & Code Quality                                        & 72&3  & 93&6   & 82&8   & 86&6  & 96&2  & 108&5   & 206&11  & 43&43  & 168&13  & 210&19  \\
 & Other&	70&4 &	87&4 &	81&14 	&90&6 	&97&2 &	105&8 &	197&26 &	45&52 &	152&16 &	190&12 \\
 & All                                                 & 70&5  & 91&0   & 82&4   & 90&1  & 97&1  & 96&6    & 197&7   & 47&8   & 142&4   & 196&7   \\
\cellcolor{white}  & \cellcolor{gray!25} Median                                              & 70&7 &	91&5 &	82&10 &	90&4 &	96&3 &	103&11 &	200&18 &	52&40 &	148&17 	&192&21   \\\hline
\multirow{7}{*}{40\%}  & Protection Mechanism Failure                                      & 55&11  & 84&3  & 67&10  & 81&4  & 89&5  & 101&13  & 205&35  & 62&50  & 149&19  & 170&57  \\
 & Resource Management Errors & 61&7   & 82&2  & 77&7   & 82&4  & 91&3  & 107&11  & 213&15  & 58&14  & 154&38  & 202&13  \\
 & Data Processing Errors                                         & 65&0   & 84&6  & 68&9   & 82&8  & 90&4  & 105&5   & 214&12  & 63&29  & 147&19  & 199&13  \\
 & Code Quality                                        & 75&3   & 86&6  & 72&6   & 82&6  & 86&6  & 108&6   & 209&17  & 45&32  & 162&16  & 208&17  \\
 &Other&	60&10 &	85&4 &	73&12 &	82&8 &	90&1 &	107&12 &	202&24 &	44&29 &	152&19 &	197&13 \\
 & All                                                 & 59&1   & 82&1  & 72&3   & 83&2  & 93&2  & 89&3    & 191&5   & 56&13  & 145&16  & 198&8   \\
\cellcolor{white} & \cellcolor{gray!25} Median                                              & 61&9 &	83&4 &	71&9 &	82&5 &	90&4 &	105&11 	&204&25 &	57&24 	&151&21 &	199&16  \\\hline
\multirow{7}{*}{50\%}  & Protection Mechanism Failure                                      & 51&5  & 76&3  & 59&18  & 70&4  & 81&5  & 112&14  & 203&27  & 58&43  & 146&41  & 184&11  \\
 & Resource Management Errors & 50&5   & 73&8  & 68&8   & 72&6  & 86&2  & 112&13  & 216&19  & 81&15  & 154&20  & 205&29  \\
 & Data Processing Errors                                         & 48&5  & 73&8  & 31&34  & 81&9  & 89&6  & 107&8  & 217&23  & 36&30  & 148&7   & 200&11  \\
 & Code Quality                                        & 51&3  & 81&9  & 62&15  & 75&5  & 87&8  & 110&7  & 219&14  & 61&27  & 166&12  & 215&12  \\
 & Other & 49&2  &  74&12 &	65&17 &	71&8 	&85&7  & 108&9 	&207&21 &	34&68 	&162&11 &	202&16  \\
 & All                                                 & 50&1  & 74&4  & 68&3   & 74&1  & 85&3  & 96&4  & 193&19  & 66&9   & 136&17  & 186&18  \\
\cellcolor{white} & \cellcolor{gray!25} Median                                              & 50&4  & 75&7 	&65&20 &	74&7 &	86&6 & 110&11 &	206&25 	&64&45 	&154&21 	&197&23  \\\hline
\end{tabular}
\begin{tablenotes}\small
This table shows the experimental results when human error rate $E_R$ increases from 0\% to 50\%. Baseline recall and cost come from Table~\ref{tab:est}, Row Text, Column 95. All the numbers are percentages and in the format of median(IQR) from 30 repeated simulations. Each row presents performances on one group of target vulnerability types in the Mozilla Firefox dataset as described in Table~\ref{tab:data}. The ``Median'' row summarizes the median performance across all groups. The ``None'' columns show the performance of {\IT} without any error correction method while other columns each reports the performance with a different error correction method. One method is considered better than another if it has higher relative recall and lower relative cost. 
\end{tablenotes}
\end{threeparttable}}
\end{center}}
\end{table*}

{\bf RQ3: Is {\IT}'s performance affected by human errors?} To show how human errors affect the performance, we take the results from {\IT} with \textbf{Text}, target recall $=$ 95\% (Table~\ref{tab:est}) as baseline result and test the same algorithm with human error rate increasing from 0\% to 50\%. Column None in Table~\ref{tab:error} shows the relative performances against the baseline result if no error correction is applied. As we can see, with 50\% human error rate, the final recall becomes half of that of baseline recall.

\begin{RQ}{RQ3: Is {\IT}'s performance affected by human errors?}
Yes. Human errors during the inspection process can adversely affect the performance of {\IT}. For example, human agents with 50\% recall result in a deterioration of recall from 96\% to 48\%.
\end{RQ}

The large deterioration in recall motivates the next  question.

{\bf RQ4: Can {\IT} correct human errors effectively?} Table~\ref{tab:error} compares DISPUTE and DISPUTE(3), the error correction methods from {\IT}, against two state-of-the-art error correction methods, Two-person and Cormack'17. One method is considered better than another if it achieves higher recall with a lower cost. From  
the last line of Table~\ref{tab:error}, we observe
that all error correction methods  achieve much higher recall than \textbf{None} (i.e. no error correction). That is, error correction is strongly recommended when human inspectors are fallible. Also, in 
 Table~\ref{tab:error}:
\bi
\item
\textbf{Cormack'17} costs least among all the error correction methods, but also achieves lower recall. Given recall as highest priority, \textbf{Cormack'17} is not recommended.
\item
\textbf{DISPUTE }is better than Two-person; i.e. it achieves similar recalls at less  cost. For example, when $E_R=50\%$,
DISPUTE reaches $0.74\times 0.96 = 71\%$ recall with $1.51\times0.26 = 39\%$ cost while Two-person reaches $0.75\times 0.96 = 72\%$ recall with $2.06\times 0.26 = 54\%$ cost.
\item
\textbf{DISPUTE(3)} is also better than Two-person; i.e. it achieves higher recall with similar cost (achieves $0.86\times 0.96 = 83\%$ recall with $1.96\times 0.26 = 51\%$ cost when $E_R=50\%$).
\item
Compared to \textbf{DISPUTE}, \textbf{DISPUTE(3)} costs more but also corrects more errors. Therefore one can always increase the number of humans rechecking the files selected by \textbf{DISPUTE} to reach higher recall with higher cost.
\item
The coverage of false negatives by \textbf{DISPUTE} is  $\frac{\mathit{Relative}\, \mathit{Recall} -(1-E_R)}{(1-E_R)E_R}=\frac{0.74-0.5}{0.25}=96\%$ when $E_R=50\%$. 
This supports the core hypothesis
of   \textbf{DISPUTE}; i.e. most human errors (96\%) are in   files selected by \textbf{DISPUTE} (50\% of the labeled files). Hence, we say:
\ei

\begin{RQ}{RQ4: Can {\IT} correct human errors effectively?}
Yes. {\IT} corrects more human errors with fewer double checks when compared to other state-of-the-art error correction methods. It double-checks 50\% of the labeled files but covers 96\% of the missing vulnerabilities.
\end{RQ}

According to our results, when targeting all types of vulnerabilities at 95\% recall with human having 50\% chance of failing to detect a vulnerability during the inspection, {\IT} with DISPUTE(3) can reach $0.95\times0.85 = 81\%$ recall for $0.21\times1.86 = 39\%$ cost, which means 81\% of the vulnerable files can be identified by inspecting 11,230 files. Although this is still a large amount of human effort, it is much better than the traditional vulnerability inspection applied in mission-critical projects without {\IT}~\cite{zelkowitz2001understanding}: 
\bi
\item
Without {\IT},
one engineer would only find 50\% of the vulnerabilities, and
only after  inspecting 28,750 files (100\% of the files); 
\item
Without {\IT},
two engineers would  find 75\% of  the vulnerabilities,
but
only
after 
inspecting 57,500 files (200\% of the files including double-checking effort).
\ei

\section{Discussion}
\label{sec:discussion}

\subsection{Limitations}
\label{sec:Limitation}

There exist several limitations in this work that we plan to resolve in the future:
\bi
\item
All the experiments in this work are conducted at
file level granularity. Currently, $Cost$ is measured as the number of files inspected. This can be improved since the effort to inspect a file can vary significantly. Also, if {\IT} can make predictions based on functions or specific lines of code instead of an entire file, it would save time and effort for the human inspector to look for potential vulnerabilities.
\item
Human errors are simulated randomly in the current work. In practice, many factors can affect the human error rate, e.g. the level of expertise of the inspector, or some certain type of vulnerabilities might be harder to find.
\item
Our simulation is based on data collected from reported security vulnerabilities and bugs. As a result, our dataset does not provide information on whether an actual incident was caused by the identified vulnerabilities. Therefore, we do not separate these two types of vulnerabilities in the simulation and we do not know what percentage of the identified vulnerabilities were involved in actual security incidents. 
\item
Low precision issue: maintaining a considerably high precision is important for such an inspection tool to be accepted and used in practice, e.g. if the human inspector finds no vulnerabilities from 10 source code files suggested by the tool, then he or she may quickly lose faith in that tool and stop using it. Although the current {\IT} tool looks promising in terms of recall and percentage of cost saved, it has low precision. As an example, in Table~\ref{tab:recall}, {\IT} reaches 95\% recall with 20\% cost when targeting all types of vulnerabilities using text mining features. This also means {\IT} finds 257 vulnerable files by asking humans to inspect 5,750 files, which results in less than 5\% precision (1 vulnerable file can be identified for every 22 files inspected). It might be acceptable in some mission-critical systems but generally speaking, users do not want to inspect 22 files only to find one vulnerability since it may take hours of careful work to inspect one source code file.
\ei

\subsection{Threats to Validity}
\label{sec:Threats to Validity}

There are several validity threats~\cite{feldt2010validity} to the design of this study. Any conclusions made from this work must be considered with the following issues in mind:

\textbf{Conclusion validity} focuses on the significance of the treatment. To enhance   conclusion validity, we ran each simulation 30 times with different random seeds.

\textbf{Internal validity }focuses on how sure we can be that the treatment caused the outcome. To enhance   internal validity, we heavily constrained our experiments to the same dataset, with the same settings, except for the treatments being compared.

\textbf{Construct validity }focuses on the relation between the theory behind the experiment and the observation. This applies to our analysis of which feature set provides the best performance. There are two concerns here.

Firstly, when we conclude that software metrics features perform worst (i.e., it contributes little to the vulnerability prediction), other reasons such as the choice of classifier (different classifiers might work better with different feature sets) might be the real cause of the observation.

Secondly, like any other data mining paper we are susceptible to biases in the labeling
of the data used to test/train this method. We have some confidence in that labeling (since two graduate students spent months of work carefully collecting those ground truth labels). Nevertheless, these two oracles could have made some systematic errors in their labeling (e.g. favoring certain kinds of vulnerabilities, and not others). To partially mitigate this problem,
we  make our data and methods available\footnote{\url{https://github.com/ai-se/Mozilla_Firefox_Vulnerability_Data}} and 
commit to helping (as requested) other research groups working on this data.

\textbf{External validity} concerns how widely our conclusions can be applied. All the conclusions in this study are drawn from the experiments running on the Mozilla Firefox vulnerability dataset. When applied to other case studies, the following concerns might arise: 1) crash dump stack trace data may not be available; 2) the same settings that work on Mozilla Firefox dataset might not provide the best performance on other datasets.  One possible solution to this problem can be hyperparameter tuning which adapts the parameter settings to the target dataset.




\section{Conclusion and Future work}\label{sec:conclusion}

Reducing software security vulnerabilities is a crucial task of software development. However, inspecting code
to find vulnerabilities is a tedious and time-consuming task. Hence, this paper introduces and evaluates
{\IT},  an active learning-based vulnerability prediction framework.  {\IT} focuses on (1) saving cost when reaching different levels of recall, (2)  providing a practical way to stop at the target recall, and (3) correcting human errors efficiently. 



This approach  was tested on a Mozilla Firefox dataset using a simulation methodology. Given  actual  vulnerabilities  reported, we run multiple “what-if” simulations where we run over that data using a variety of code inspection strategies. The goal of  these  inspections  is  to  determine  what  might  have  happened if  these  strategies  were  applied for finding vulnerabilities before the software is released.

What we found was that using text mining features alone, {\IT} decreases the cost to reach high recall for finding vulnerabilities before deployment, and if runtime data is available (e.g. the crash features used above), then that can boost vulnerability prediction in the early stages. We also showed that the total number of vulnerabilities in one software project can be  accurately estimated during the active learning process, thus providing a reliable stopping rule for the approach. Based on our results,  {\IT} can save 77, 70, 53\% of the cost (compared to reviewing and testing source code files in a random order) when applying the SEMI estimator to stop at 90, 95, 99\% recall, respectively. Note that these results indicate that {\IT}  can result in, by practitioner choice, a recall value close to 100\% given trade-offs in recall and cost. Practitioners can have confidence that, if they so choose and have the resources to spare, 99\% of the vulnerabilities would be identified without inspecting most of the source code. Meanwhile, {\IT} can cover 96\% of the human missing vulnerable files by double-checking 50\% of the inspected files, thus saving more effort when correcting human errors. 

\begin{table}[!t]
\caption{Comparison to the state-of-the-art frameworks}
\label{tab:compare}
\begin{center}
\setlength\tabcolsep{5pt}
\begin{tabular}{r|ccc}
& {\IT} &  \makecell{Supervised or \\Semi-supervised} & \makecell{Unsupervised}
\\\hline
\makecell[r]{Can reach any \\target recall} & \checkmark & \checkmark & \\\hline
\makecell[r]{Guide on how to \\reach target recall} & \checkmark &  & \\\hline
\makecell[r]{Can start without \\known vulnerabilities} & \checkmark &  & \checkmark \\\hline
\makecell[r]{Work without \\other indicators} & \checkmark & \checkmark &\\\hline
\makecell[r]{Utilize continuous \\human feedback} & \checkmark &  & \\\hline
\makecell[r]{Correct human \\errors efficiently} & \checkmark &  & \\
\hline
\end{tabular}
\end{center}
\end{table}

Table~\ref{tab:compare} compares the advantages of {\IT} with other approaches to vulnerability prediction, i.e. supervised or semi-supervised approaches~\cite{shin2011evaluating,zimmermann2010searching,7853039} and unsupervised approaches~\cite{theisen2015strengthening,Theisen2015ICSE,theisen2017risk}.  We assert that {\IT} does best due to (a) the capability to make full use of  continuous human feedback, (b) the guidance of when to stop inspections, and (c) the human error correction strategy.

That said, {\IT} still suffers from the low precision problem as discussed in \tion{Limitation}. Even after the reduction, the inspection effort required is still very high (inspecting 5,750 files in our case study). Therefore much research is required to further reduce the effort
associated with vulnerability inspection.
There are two possible directions for further cost reduction: (a) increasing the precision by improving the model's prediction, and (b) reducing the cost of every false alarm.

As for improving the model's prediction:
\bi
\item
Combining text mining features with other vulnerability-proneness metrics might improve the model's performance as suggested by the \textbf{Hybrid} results.
\item
Although our current way of using semi-supervised learning did not produce better results, it is still possible to improve {\IT}'s performance by better utilizing the information from unlabeled data. For example,
semi-supervised learning could become a pre-filter
that removes from consideration the files
that are less likely to contain vulnerabilities. 
\item
This work applies some best practice settings (featurization, classifier, active learning, etc.) from other domains, this does not necessarily be the best practice for proactive vulnerability prediction. Further testing of other settings and tuning of the parameters might provide improved results on this specific domain.
\ei

As for reducing the cost of every false alarm:
\bi
\item
One direction is to predict based on function level or lines of code level, so that the cost of every false alarm becomes the human effort of inspecting one line of code instead of one file.
\item
The other direction is to facilitate the inspection by highlighting which part of the code in the file contains terms used by the prediction model.
\ei

Apart from the low precision issue, considering the limitations and validity threats of the current work, other future works could include:
\begin{enumerate}
\item
The community requires more dataset like the Mozilla Firefox vulnerability dataset we described in this paper. Hence we encourage software engineering researchers to collect more high-quality vulnerability datasets and test the generalizability of vulnerability prediction algorithms.
\item
Solicit opinions on the results from the development team that would take action based on the results. In our study, we reached out to contacts at the Mozilla Foundation, but did not get a response. In the future, field research should be conducted with real human inspectors using {\IT}.
\item
As shown in \tion{errors}, different level of human error rate $E_R$ requires different error correction methods. E.g. if $E_R=0$, using DISPUTE will cost 46\% more than not using any error correction; and at high error rate $E_R=50\%$, DISPUTE(3) is more desired than DISPUTE. Therefore how to accurately estimate or measure the human error rate and adjust the error correction method correspondingly becomes an important future work.
\item
Check generality of the active learning framework by applying it to other total recall problems in software engineering. The authors are already exploring SE problems like test case prioritization, technical debt detection, and static warning identification. Preliminary results suggest that the active learning framework discussed here might be widely applicable to many SE problems.
\end{enumerate}
We hope that this work will lay down a foundation and motivate further research on active learning-based vulnerability prediction.

\section*{Acknowledgements}\label{sec:acks}

This research was partially funded by a National Science Foundation Grant \#1506586
and \#1909516. The authors would like to thank the Mozilla Firefox team for making crash and defect data available to the public, which makes this research possible.

\bibliographystyle{IEEEtran}

\bibliography{mybib}

\newpage

\begin{IEEEbiography}[{\includegraphics[width=1in,clip,keepaspectratio]{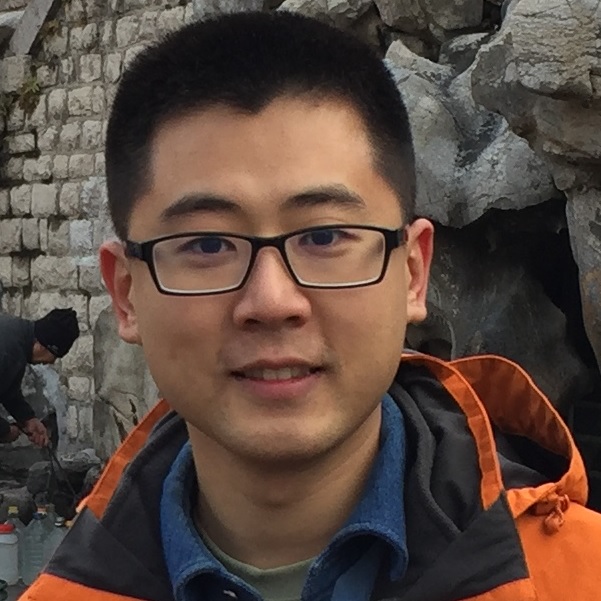}}]{Zhe Yu}
 is a fifth year Ph.D. student in Computer Science at North Carolina State University. He received his masters degree in Shanghai Jiao Tong University, China. His primary interest lies in the collaboration of human and machine learning algorithms that leads to better performance and higher efficiency. \url{http://azhe825.github.io/}
\end{IEEEbiography}
\begin{IEEEbiography}[{\includegraphics[width=1in,clip,keepaspectratio]{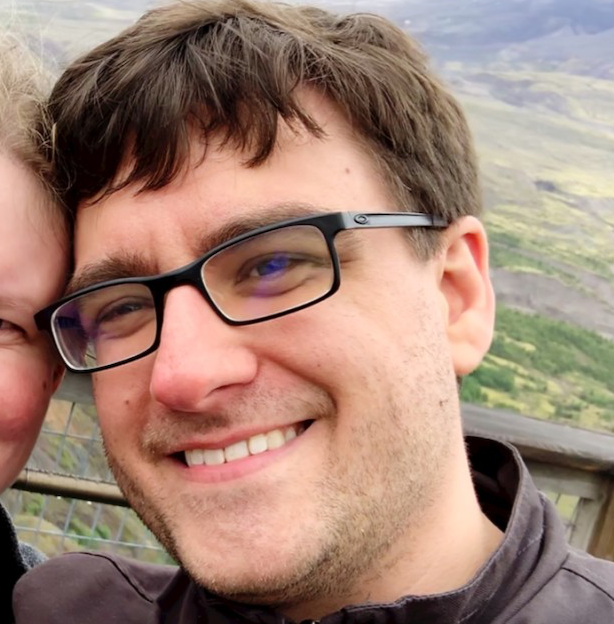}}]{Christopher Theisen}
is currently a Program Manager at Microsoft. He received his Ph.D. in Computer Science from North Carolina State University. His research is focused in software security and software engineering, specifically developing attack surface metrics. \url{http://theisencr.github.io/}
\end{IEEEbiography}
\begin{IEEEbiography}[{\includegraphics[width=1in,clip,keepaspectratio]{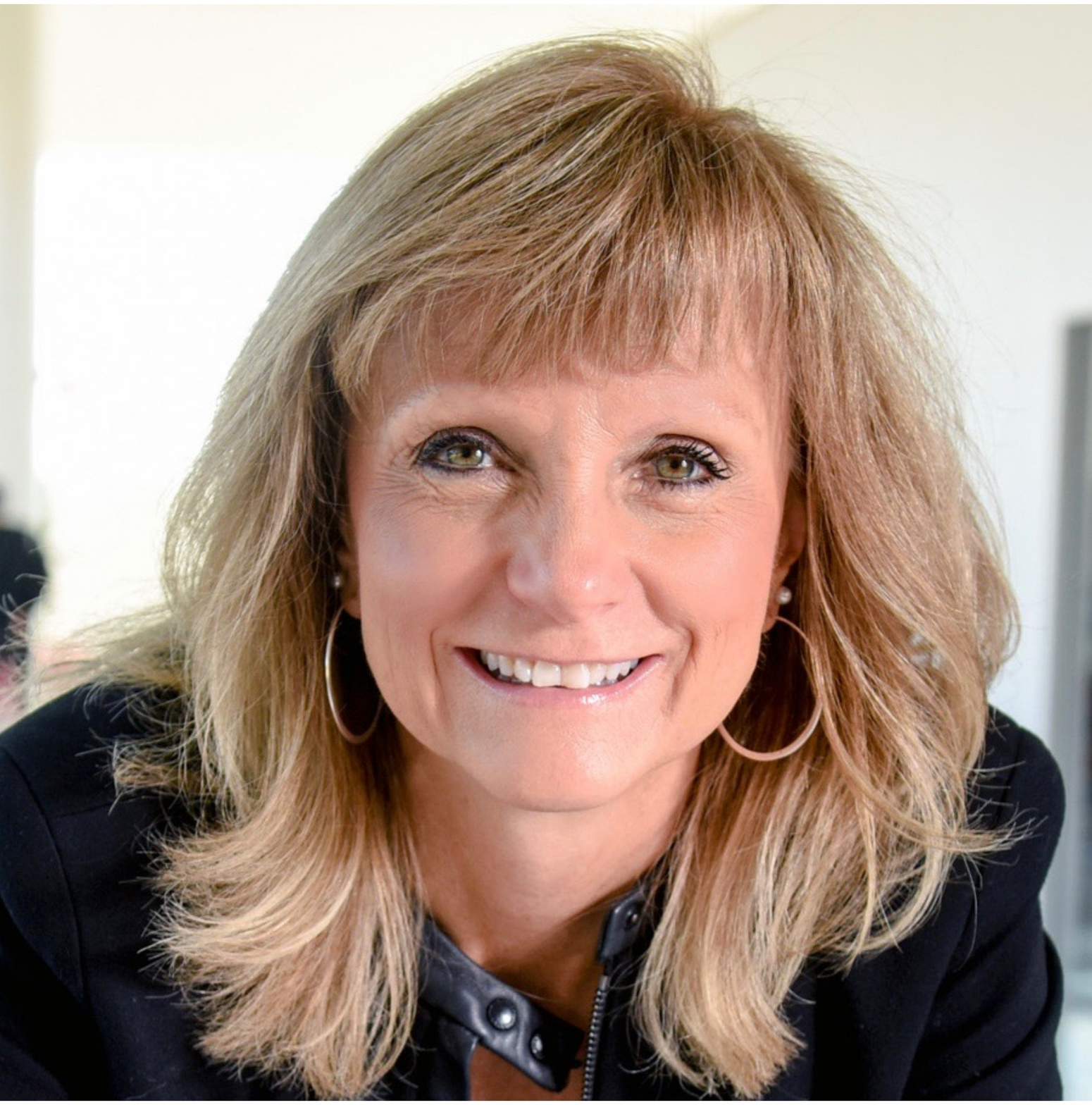}}]{Laurie Williams} (IEEE Fellow)
 received her Ph.D. in Computer Science from the University of Utah.  Her research focuses on software security; agile software development practices and processes, particularly continuous deployment; and software reliability, software testing and analysis. Prof. Williams has more than 230 refereed publications.   \url{https://collaboration.csc.ncsu.edu/laurie/}
\end{IEEEbiography}
\begin{IEEEbiography}[{\includegraphics[width=1in,clip,keepaspectratio]{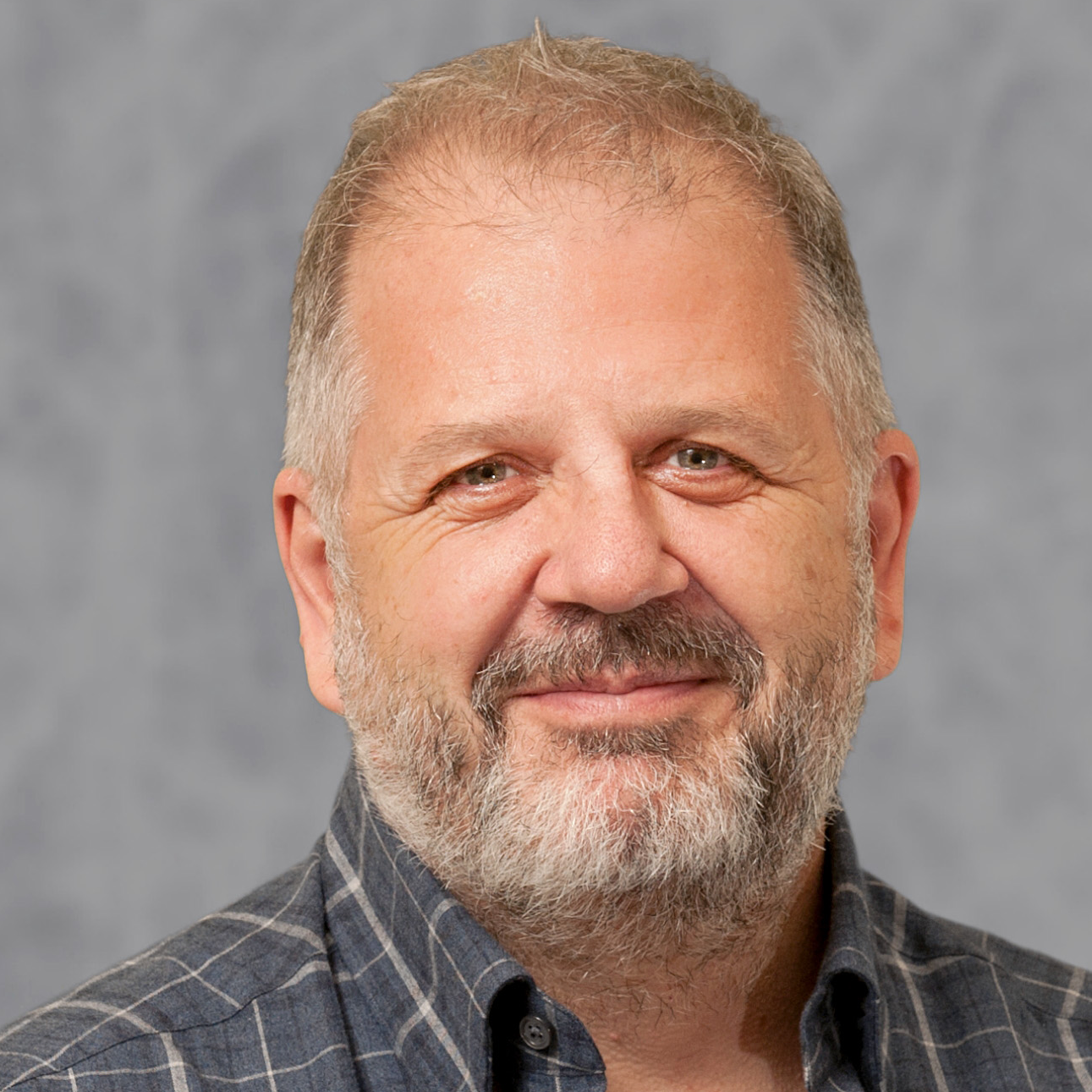}}]{Tim Menzies} (IEEE Fellow)
is a Professor in CS at NcState  His research interests include software engineering (SE), data mining, artificial intelligence, search-based SE, and open access science. \url{http://menzies.us}
\end{IEEEbiography}



\end{document}